\documentclass[twocolumn]{aastex631}

\newcommand\spinny{\texttt{SPINNY}}
\newcommand\multimoon{\texttt{MultiMoon}}
\newcommand\jt{$J_2$}
\newcommand\ct{$C_{22}$}
\newcommand\kagara{\textdoublebarpipe K\cb{\'{a}}g\'{a}ra-!H\~{a}unu}

\newcommand{\trackchange}[1]{\textit{{\color{red} #1}}}

\usepackage[utf8]{inputenc}
\usepackage{tipa}
\usepackage{combelow}
\usepackage{savesym}
\savesymbol{tablenum}
\usepackage{siunitx}
\restoresymbol{SIX}{tablenum}
\usepackage{breqn}
\usepackage{tabularx}
\usepackage{placeins}

\shorttitle{TNO Binary Shapes}
\shortauthors{Proudfoot et al.}

\begin{document}

\title{Beyond Point Masses. II. Non-Keplerian Shape Effects are Detectable in Several TNO Binaries}

\correspondingauthor{Benjamin Proudfoot}
\email{benp175@gmail.com}

\author[0000-0002-1788-870X]{Benjamin C.N. Proudfoot}
\affiliation{Brigham Young University Department of Physics \& Astronomy, N283 ESC, Brigham Young University, Provo, UT 84602, USA}
\affiliation{Florida Space Institute, University of Central Florida, 12354 Research Parkway, Orlando, FL 32826, USA}
    
\author[0000-0003-1080-9770]{Darin A. Ragozzine}
\affiliation{Brigham Young University Department of Physics \& Astronomy, N283 ESC, Brigham Young University, Provo, UT 84602, USA}
    
\author[0000-0001-6921-7070]{Meagan L. Thatcher}
\affiliation{Brigham Young University Department of Physics \& Astronomy, N283 ESC, Brigham Young University, Provo, UT 84602, USA}
    
\author[0000-0002-8296-6540]{Will Grundy}
\affiliation{Lowell Observatory, Flagstaff, AZ, USA}

\author[0000-0003-4051-2003]{Dallin J. Spencer}
\affiliation{Brigham Young University Department of Physics \& Astronomy, N283 ESC, Brigham Young University, Provo, UT 84602, USA}
    
\author{Tahina M. Alailima}
\affiliation{Brigham Young University Department of Physics \& Astronomy, N283 ESC, Brigham Young University, Provo, UT 84602, USA}
\author{Sawyer Allen}
\affiliation{Brigham Young University Department of Physics \& Astronomy, N283 ESC, Brigham Young University, Provo, UT 84602, USA}
\author{Penelope C. Bowden}
\affiliation{Brigham Young University Department of Physics \& Astronomy, N283 ESC, Brigham Young University, Provo, UT 84602, USA}
\author{Susanne Byrd}
\affiliation{Brigham Young University Department of Physics \& Astronomy, N283 ESC, Brigham Young University, Provo, UT 84602, USA}
\author[0000-0002-8784-5780]{Conner D. Camacho}
\affiliation{Brigham Young University Department of Physics \& Astronomy, N283 ESC, Brigham Young University, Provo, UT 84602, USA}
\author{Gibson H. Campbell}
\affiliation{Brigham Young University Department of Physics \& Astronomy, N283 ESC, Brigham Young University, Provo, UT 84602, USA}
\author{Edison P. Carlisle}
\affiliation{Brigham Young University Department of Physics \& Astronomy, N283 ESC, Brigham Young University, Provo, UT 84602, USA}
\author{Jacob A. Christensen}
\affiliation{Brigham Young University Department of Physics \& Astronomy, N283 ESC, Brigham Young University, Provo, UT 84602, USA}
\author{Noah K. Christensen}
\affiliation{Brigham Young University Department of Physics \& Astronomy, N283 ESC, Brigham Young University, Provo, UT 84602, USA}
\author{Kaelyn Clement}
\affiliation{Brigham Young University Department of Physics \& Astronomy, N283 ESC, Brigham Young University, Provo, UT 84602, USA}
\author{Benjamin J. Derieg}
\affiliation{Brigham Young University Department of Physics \& Astronomy, N283 ESC, Brigham Young University, Provo, UT 84602, USA}
\author{Mara K. Dille}
\affiliation{Brigham Young University Department of Physics \& Astronomy, N283 ESC, Brigham Young University, Provo, UT 84602, USA}
\author{Cristian Dorrett}
\affiliation{Brigham Young University Department of Physics \& Astronomy, N283 ESC, Brigham Young University, Provo, UT 84602, USA}
\author{Abigail L. Ellefson}
\affiliation{Brigham Young University Department of Physics \& Astronomy, N283 ESC, Brigham Young University, Provo, UT 84602, USA}
\author{Taylor S. Fleming}
\affiliation{Brigham Young University Department of Physics \& Astronomy, N283 ESC, Brigham Young University, Provo, UT 84602, USA}
\author{N. J. Freeman}
\affiliation{Brigham Young University Department of Physics \& Astronomy, N283 ESC, Brigham Young University, Provo, UT 84602, USA}
\author{Ethan J. Gibson}
\affiliation{Brigham Young University Department of Physics \& Astronomy, N283 ESC, Brigham Young University, Provo, UT 84602, USA}
\author[0000-0001-6838-1530]{William G. Giforos}
\affiliation{Brigham Young University Department of Physics \& Astronomy, N283 ESC, Brigham Young University, Provo, UT 84602, USA}
\author{Jacob A. Guerrette}
\affiliation{Brigham Young University Department of Physics \& Astronomy, N283 ESC, Brigham Young University, Provo, UT 84602, USA}
\author{Olivia Haddock}
\affiliation{Brigham Young University Department of Physics \& Astronomy, N283 ESC, Brigham Young University, Provo, UT 84602, USA}
\author{S. Ashton Hammond}
\affiliation{Brigham Young University Department of Physics \& Astronomy, N283 ESC, Brigham Young University, Provo, UT 84602, USA}
\author{Zachary A. Hampson}
\affiliation{Brigham Young University Department of Physics \& Astronomy, N283 ESC, Brigham Young University, Provo, UT 84602, USA}
\author{Joshua D. Hancock}
\affiliation{Brigham Young University Department of Physics \& Astronomy, N283 ESC, Brigham Young University, Provo, UT 84602, USA}
\author{Madeline S. Harmer}
\affiliation{Brigham Young University Department of Physics \& Astronomy, N283 ESC, Brigham Young University, Provo, UT 84602, USA}
\author{Joseph R. Henderson}
\affiliation{Brigham Young University Department of Physics \& Astronomy, N283 ESC, Brigham Young University, Provo, UT 84602, USA}
\author{Chandler R. Jensen}
\affiliation{Brigham Young University Department of Physics \& Astronomy, N283 ESC, Brigham Young University, Provo, UT 84602, USA}
\author{David Jensen}
\affiliation{Brigham Young University Department of Physics \& Astronomy, N283 ESC, Brigham Young University, Provo, UT 84602, USA}
\author{Ryleigh E. Jensen}
\affiliation{Brigham Young University Department of Physics \& Astronomy, N283 ESC, Brigham Young University, Provo, UT 84602, USA}
\author{Joshua S. Jones}
\affiliation{Brigham Young University Department of Physics \& Astronomy, N283 ESC, Brigham Young University, Provo, UT 84602, USA}
\author{Cameron C. Kubal}
\affiliation{Brigham Young University Department of Physics \& Astronomy, N283 ESC, Brigham Young University, Provo, UT 84602, USA}
\author{Jacob N. Lunt}
\affiliation{Brigham Young University Department of Physics \& Astronomy, N283 ESC, Brigham Young University, Provo, UT 84602, USA}
\author{Stephanie Martins}
\affiliation{Brigham Young University Department of Physics \& Astronomy, N283 ESC, Brigham Young University, Provo, UT 84602, USA}
\author{McKenna Matheson}
\affiliation{Brigham Young University Department of Physics \& Astronomy, N283 ESC, Brigham Young University, Provo, UT 84602, USA}
\author{Dahlia Maxwell}
\affiliation{Brigham Young University Department of Physics \& Astronomy, N283 ESC, Brigham Young University, Provo, UT 84602, USA}
\author{Timothy D. Morrell}
\affiliation{Brigham Young University Department of Physics \& Astronomy, N283 ESC, Brigham Young University, Provo, UT 84602, USA}
\author{McKenna M. Myckowiak}
\affiliation{Brigham Young University Department of Physics \& Astronomy, N283 ESC, Brigham Young University, Provo, UT 84602, USA}
\author[0000-0001-8780-8480]{Maia A. Nelsen}
\affiliation{Brigham Young University Department of Physics \& Astronomy, N283 ESC, Brigham Young University, Provo, UT 84602, USA}
\author{Spencer T. Neu}
\affiliation{Brigham Young University Department of Physics \& Astronomy, N283 ESC, Brigham Young University, Provo, UT 84602, USA}
\author{Giovanna G. Nuccitelli}
\affiliation{Brigham Young University Department of Physics \& Astronomy, N283 ESC, Brigham Young University, Provo, UT 84602, USA}
\author{Kayson M. Reardon }
\affiliation{Brigham Young University Department of Physics \& Astronomy, N283 ESC, Brigham Young University, Provo, UT 84602, USA}
\author{Austin S. Reid}
\affiliation{Brigham Young University Department of Physics \& Astronomy, N283 ESC, Brigham Young University, Provo, UT 84602, USA}
\author{Kenneth G. Richards }
\affiliation{Brigham Young University Department of Physics \& Astronomy, N283 ESC, Brigham Young University, Provo, UT 84602, USA}
\author{Megan R. W. Robertson}
\affiliation{Brigham Young University Department of Physics \& Astronomy, N283 ESC, Brigham Young University, Provo, UT 84602, USA}
\author{Tanner D. Rydalch}
\affiliation{Brigham Young University Department of Physics \& Astronomy, N283 ESC, Brigham Young University, Provo, UT 84602, USA}
\author{Conner B. Scoresby}
\affiliation{Brigham Young University Department of Physics \& Astronomy, N283 ESC, Brigham Young University, Provo, UT 84602, USA}
\author{Ryan L. Scott }
\affiliation{Brigham Young University Department of Physics \& Astronomy, N283 ESC, Brigham Young University, Provo, UT 84602, USA}
\author[0000-0002-1286-061X]{Zacory D. Shakespear}
\affiliation{Brigham Young University Department of Physics \& Astronomy, N283 ESC, Brigham Young University, Provo, UT 84602, USA}
\author{Elliot A.Silveira }
\affiliation{Brigham Young University Department of Physics \& Astronomy, N283 ESC, Brigham Young University, Provo, UT 84602, USA}
\author{Grace C. Steed }
\affiliation{Brigham Young University Department of Physics \& Astronomy, N283 ESC, Brigham Young University, Provo, UT 84602, USA}
\author{Christiana Z. Suggs}
\affiliation{Brigham Young University Department of Physics \& Astronomy, N283 ESC, Brigham Young University, Provo, UT 84602, USA}
\author{Garrett D. Suggs}
\affiliation{Brigham Young University Department of Physics \& Astronomy, N283 ESC, Brigham Young University, Provo, UT 84602, USA}
\author{Derek M. Tobias}
\affiliation{Brigham Young University Department of Physics \& Astronomy, N283 ESC, Brigham Young University, Provo, UT 84602, USA}
\author{Matthew L. Toole}
\affiliation{Brigham Young University Department of Physics \& Astronomy, N283 ESC, Brigham Young University, Provo, UT 84602, USA}
\author{McKayla L. Townsend }
\affiliation{Brigham Young University Department of Physics \& Astronomy, N283 ESC, Brigham Young University, Provo, UT 84602, USA}
\author{Kade L. Vickers}
\affiliation{Brigham Young University Department of Physics \& Astronomy, N283 ESC, Brigham Young University, Provo, UT 84602, USA}
\author{Collin R. Wagner}
\affiliation{Brigham Young University Department of Physics \& Astronomy, N283 ESC, Brigham Young University, Provo, UT 84602, USA}
\author{Madeline S. Wright}
\affiliation{Brigham Young University Department of Physics \& Astronomy, N283 ESC, Brigham Young University, Provo, UT 84602, USA}
\author{Emma M. A. Zappala}
\affiliation{Brigham Young University Department of Physics \& Astronomy, N283 ESC, Brigham Young University, Provo, UT 84602, USA}

\begin{abstract} 
About 40 transneptunian binaries (TNBs) have fully determined orbits with about 10 others being solved except for breaking the mirror ambiguity. Despite decades of study almost all TNBs have only ever been analyzed with a model that assumes perfect Keplerian motion (e.g., two point masses). In reality, all TNB systems are non-Keplerian due to non-spherical shapes, possible presence of undetected system components, and/or solar perturbations. In this work, we focus on identifying candidates for detectable non-Keplerian motion based on sample of 45 well-characterized binaries. We use \multimoon, a non-Keplerian Bayesian inference tool, to analyze published relative astrometry allowing for non-spherical shapes of each TNB system's primary. We first reproduce the results of previous Keplerian fitting efforts with \multimoon, which serves as a comparison for the non-Keplerian fits and confirms that these fits are not biased by the assumption of a Keplerian orbit. We unambiguously detect non-Keplerian motion in 8 TNB systems across a range of primary radii, mutual orbit separations, and system masses. As a proof of concept for non-Keplerian fitting, we perform detailed fits for (66652) Borasisi-Pabu, possibly revealing a $J_2 \approx 0.44$, implying Borasisi (and/or Pabu) may be a contact binary or an unresolved compact binary. However, full confirmation of this result will require new observations. This work begins the next generation of TNB analyses that go beyond the point mass assumption to provide unique and valuable information on the physical properties of TNBs with implications for their formation and evolution.
\end{abstract}

\section{Introduction}

Since the discovery and characterization of the mutual binary orbit of the transneptunian object (TNO) 1998 WW31 \citep{2002Natur.416..711V}, transneptunian binaries (TNBs) have been recognized as sensitive tracers of the history of the solar system. Acting as a detailed laboratory that enables mass measurements, TNBs open the door for remote characterization of TNOs as a whole \citep[e.g.,][]{2007Icar..191..286G,grundy2009mutual,fraser2010quaoar,2016MNRAS.460.1542B}. In addition, the mutual orbital properties of a binary system provide insight into the formation and history of that binary system, as those properties encode information about the binary's formation \citep[e.g.,][]{2007Sci...316.1585B, brown2010size}, past tidal evolution \citep[e.g.,][]{porter2012kctf,arakawa2021tidal}, collisional history \citep[e.g.,][]{2009AJ....137.4766R,parker2011collisional}, and encounters with other bodies \citep[e.g.,][]{campbell2022close}. The statistical ensemble of mutual orbit properties of TNBs also hold valuable clues about the conditions of the protoplanetary disk from which TNBs originally formed and have revealed groundbreaking insights into the dominant formation processes in that disk \citep[e.g.,][]{nesvorny2010formation,grundy2019mutual,nesvorny2019trans}. 

While the numerous studies of TNBs have enabled unprecedented understanding of processes in the outer solar system, it has become clear that current state-of-the-art methods are lagging behind the growing observational baselines of TNBs. These methods, for the most part, rely on simple Keplerian orbital models, with only a few exceptions. In recent years, however, many authors have found that the observed relative astrometric positions of TNBs show statistically significant deviations from plain Keplerian orbits (Salacia-Actaea 3.7$\sigma$, Orcus-Vanth 2.2$\sigma$, \citet{grundy2019mutual}; Eris-Dysnomia 6.3$\sigma$, \citet{holler2021eris}). An analysis of the distribution of $\chi^2$ values of all TNB fits also shows a statistically significant excess of poor fits. While these deviations could be the result of unidentified systematic errors in data collections/measurement, it is also likely that the deviations are the manifestation of inaccurate models including non-Keplerian gravitational effects acting in TNB systems. 

Non-Keplerian gravitational effects are any gravitational effect that forces an orbit to deviate from a pure Keplerian orbit. Generally, non-Keplerian effects result in slow precession of an orbit's orientation angles. Precession of the direction of periapse is called apsidal precession, and precession of the orbit pole is called nodal precession. While there are many possible sources of non-Keplerian effects, the most relevant sources for TNBs are the non-spherical shapes of individual TNB components, systems with more than 2 components (whether known or unknown), and the external gravitational influence of the Sun (see \citet{bpm1spinnymultimoon} for more details). 

In this paper, we will focus on detecting and measuring the strength of non-Keplerian effects attributable to shape or unknown components, leaving non-Keplerian effects from solar influences to future study. In Section \ref{sec:shapes}, we discuss the causes and consequences of non-Keplerian shape effects. Next, in Section \ref{sec:orbitfitting}, we explain the general process of our non-Keplerian orbit fitting. In Section \ref{sec:kep_methods} and Section \ref{sec:nonkep_methods}, we discuss our methods for both Keplerian and non-Keplerian orbital fitting applied herein. In Section \ref{sec:data}, we detail our TNB sample and gather publicly available relative astrometric data. In Section \ref{sec:keplerianfits}, we discuss the results of our Keplerian orbit fits and reproduce past fitting results. Next, in Section \ref{sec:nonkeplerian_fits}, we present the results of our search for non-Keplerian effects, discuss the implications of these results, and identify the most promising targets for future investigation. Then, in Section \ref{sec:borasisi}, we perform a full non-Keplerian analysis of TNB (66652) Borasisi-Pabu, one of the most promising cases of non-Keplerian motion in a TNB, as a proof-of-concept of full non-Keplerian fitting. Lastly, in Section \ref{sec:conclusion}, we discuss our conclusions. 

\subsection{Non-Keplerian Shape Effects}
\label{sec:shapes}

Keplerian orbital models implicitly assume that the individual components of a TNB system are point masses (or equivalently perfect spheres). However, it is well-documented that the shapes of TNOs can be significantly non-spherical \citep[e.g.,][]{sheppard2004extreme,ortiz2017size}. Moving beyond point masses, non-spherical shapes must cause non-Keplerian deviations in TNB orbits, though the importance of these deviations relative to present or future observational data has not yet been previously examined in detail. 

Gravitational potentials of non-spherical bodies can be modeled using a spherical harmonic expansion of the gravitational potential. Current data warrant exploring the gravitational potential only at second order; by construction, higher order corrections are less important and neglected in our analysis. \citet{bpm1spinnymultimoon} provide a detailed discussion of these and other effects and how they can be modeled which we summarize here. 

The second-order gravitational potential, $U$, of a mass $M$ at distance $r$, can be written:

\begin{dmath}\label{eqn:potential}
    U(r,\theta,\phi) = -\frac{GM}{r} \left[
    1 - J_2 \left( \frac{R}{r} \right) ^2 \left( \frac{3}{2}\sin^2\theta - \frac{1}{2} \right) 
    + C_{22} \left( \frac{R}{r} \right) ^2\cos^2\theta\sin2\phi + 
    \mathcal{O} \left( r^{-3} \right) \right]
\end{dmath}

\noindent where \jt{} is the second-order zonal gravitational harmonic and \ct{} is the second-order sectoral gravitational harmonic coefficient, $\theta$ is the body-fixed latitude-like angle, $\phi$ is the body-fixed longitude-like angle (chosen to eliminate other terms), and $R$ is a reference radius \citep[][]{1995geph.conf....1Y, scheeres2000evaluation}. \jt{} is a measure of the oblateness of the potential and \ct{} is related to the prolateness, or the ellipticity of the equator. Generally, for extremely spherical bodies (like the Earth) \jt{} $\lesssim$ 0.001, extremely oblate bodies (like Haumea) \jt{} $\sim$ 0.1, and contact binaries have \jt{} $\sim$ 0.3. 



\noindent For most TNBs, where the TNBs mutual orbit is much slower than the rotations of the individual components, \ct{} has little effect on the dynamics of the mutual orbit, as the contribution from \ct{} averages out. However, near spin-orbit resonances, which may be present among some TNB systems, \ct{} may not fully average out and may play a significant role in the dynamics of the system \citep[][]{proudfoot2021prolate}. We consider this to be a special case not relevant to most TNBs. Since we are exploring the entire ensemble of TNBs, we focus this analyses on \jt{} alone. Implications for this choice are discussed further in Section \ref{sec:borasisi}.  

The dynamics of a system with a \jt{} can be described by slow apsidal and nodal precession. For a test particle orbiting around a body with a \jt, the apsidal and nodal secular rates can be written:

\begin{equation}
\label{eqn:apsidal}
    \dot{\omega} = -\frac{3nJ_2R^2}{2a^2(1-e^2)^2}\left( \frac{5}{2}\sin^2i-2\right)
\end{equation}
\begin{equation}
\label{eqn:nodal}
    \dot{\Omega} = -\frac{3nJ_2R^2}{2a^2(1-e^2)^2}\cos{i}
\end{equation}
\noindent where $n$ is the mean motion, $a$ is the semi-major axis, $e$ is the eccentricity, and $i$ is the inclination of the orbit relative to the body's equator. As discussed in \citet{2009AJ....137.4766R}, and is clear from Equations \ref{eqn:apsidal} and \ref{eqn:nodal}, the non-Keplerian motion induced by the gravitational harmonics require knowledge of both the strength of the gravitational harmonics \textit{and} the direction of the spin axis (to determine the appropriate $\theta$ and $\phi$ values in Equation \ref{eqn:potential} or to determine the appropriate inclination in equations \ref{eqn:apsidal} and \ref{eqn:nodal}). Thus, detection of non-Keplerian effects allows constraints to be placed on both TNO shapes and spin poles (assumed to be identical to the quadrupolar gravitational harmonic pole appropriate for TNOs which are too large to sustain long-term non-principal axis rotation).

From Equations \ref{eqn:potential}, \ref{eqn:apsidal}, and \ref{eqn:nodal}, it can be seen that the measurable quantity is $J_2R^2$ (and $C_{22}R^2$) and not $J_2$ independently. This is extremely similar to the quantity $GM$, where the gravitational constant and mass are always paired together. The interpretation of $J_2R^2$ in terms of an object's shape requires first the choice of a shape model. For example, assuming a homogeneous oblate ellipsoid with semi-axes $a = b \ge c$ gives $\frac{a}{c} = \frac{b}{c} = \sqrt{5\frac{J_2R^2}{c^2} + 1}$. Different shape models can match the observed $J_2R^2$ even given a specific mass \citep[see, e.g., ][]{marchis2005mass, bpm1spinnymultimoon} and the shape model influences the choice of $R$ and thus $J_2$. Unknown contact and compact binaries can also result in an apparent $J_2R^2$, as discussed below. Except for the largest TNOs, $J_2R^2$ is dominated by the overall shape of the object and not the interior density distribution. 

However, the inability to determine a precise shape or interior composition with $J_2R^2$ should not detract from the fact that measuring a significant $J_2R^2$ has important implications for the object. For example, different formation modes can result in different $J_2R^2$ values with catastrophic collisions reaccumulating into nearly spherical objects while high angular momentum formation processes tending to form more non-spherical shapes. Of course, the strength of $J_2R^2$ affects long-term spin-orbit-tidal dynamics and evolution of the TNBs themselves \citep[e.g.,][]{correia2018chaotic}. The measurement of $J_2R^2$ also provides mostly orthogonal constraints on shape properties, so that it improves the interpretation of light curves, occultations, and thermal measurements. 

The rotation poles of TNOs have also proved to be difficult to constrain, although some success has been seen among the Centaurs \citep[e.g.,][]{tegler2005period,duffard2014photometric}. Until now, determination of shape has relied on both light curve and occultation studies. While light curves are powerful tools for understanding shape and spin, the shape and spin solutions produced are non-unique \citep[][]{harris2020asteroid}. Additionally, light curve inversion requires observations of the target at a variety of aspects, which is implausible for most TNOs due to their centuries-long heliocentric orbital periods. On the other hand, occultations can directly observe the shape of a TNO in a projected plane \citep[e.g.,][]{elliot2010size,benedetti2016results,ortiz2017size}, placing limits on the full 3-dimensional shape of a body when combined with a light curve. However, due to the random timing of the events and difficulty of observation, occultations cannot yet be used to systematically understand the shapes and spins of TNOs at a population level. 

Detection and measurement of non-Keplerian effects thus opens the door to a deeper understanding of TNB systems. While Keplerian orbital fitting allows for determination of mass (and therefore density), non-Keplerian orbit fitting in theory allows for study of the three dimensional shape and spin orientation of the two system components. It is able to provide unique constraints on both the shapes of TNB components and their obliquities\footnote{Throughout this paper, obliquity will refer to the inclination of the mutual orbit with respect to the primary body's equator.} \citep[][]{2009AJ....137.4766R,vachier2012determination} which are sensitive tracers of formation and evolution \citep[e.g.,][]{mckinnon2020solar}. 

\subsubsection{Unknown Components}

Another area where non-Keplerian analysis has significant potential for unique insight is in the detection of unknown components whether in ``contact binaries'' or simply closely separated objects (``compact binaries''). Many channels of evidence suggest that these are common in the TNO population, including light curve studies \citep[][]{rabinowitz2019complex, thirouin2018plutino, thirouin2019light,showalter2021statistical}, occultations \citep[][]{leiva2020stellar}, and imaging from within the TNO population as enabled by the \textit{New Horizons} spacecraft \citep[][]{weaver2022high}. Indeed, the only small TNO ever visited by a spacecraft, Arrokoth, was a contact binary \citep[][]{stern2019initial,mckinnon2020solar}. Discovering and characterizing contact or tight components in binaries provides unique insight into the angular momentum and mass budget of TNBs.

There is a huge range of parameter space where additional components would be too far away to be seen in light curves or occultations ($\gtrsim$3 primary radii away) but too close to be resolved in direct imaging ($\lesssim$20 primary radii away for most TNBs). Non-Keplerian effects, on the other hand, are highly enhanced in both contact and unresolved binaries.

Both non-spherical shapes and additional components primarily cause orbital precession. Indeed, to quadrupole order, the gravitational potential of a close-in, previously undetected satellite is effectively the same as unusually large gravitational harmonics. This means that exploration of non-Keplerian effects using a \jt{} shape model is a good starting approximation for detecting additional components. Detecting anomalously large values for \jt{} (e.g., $J_2 \gtrsim 1$ when assuming a triaxial shape model), may indicate that there are additional undetected system components.  

Using the definitions of the gravitational harmonics from \citet{1995geph.conf....1Y}, we find that the `effective' \jt{} and \ct{} of a close-in satellite would be
\begin{equation}\label{eqn:j2binary}
    J_{2}R^2 = \frac{1}{2} \frac{q}{\left(1+q\right)^2} a_s^2
\end{equation}
\begin{equation}\label{eqn:c22binary}
    C_{22}R^2 = \frac{1}{4} \frac{q}{\left(1+q\right)^2} a_s^2
\end{equation}
\noindent where $q = m_s/m_p$, $m_p$ and $m_s$ are the masses of the masses of the primary and (unresolved) secondary, respectively, and $a_s$ is the semi-major axis of the secondary's orbit. In this equation, we leave the combinations $J_{2}R^2$ and $C_{22}R^2$ since these are the physically meaningful parameters (see Equation \ref{eqn:potential}). It is possible to understand these contact/compact binaries as one end of the continuum of shape modeling that converts the measurement of $J_2R^2$ into a physical configuration of mass. 

These contact and compact binaries describe one (or both) of the components of an existing TNB, converting binaries into hierarchical triples (or quadruples). These systems, only one of which was previously known \citep[Lempo;][]{benecchi201047171}, are common outcomes of gravitational collapse simulations \citep[][]{nesvorny2010formation, robinson2020investigating, nesvorny2021binary}, although they may also be able to form through capture mechanisms \citep[][]{brunini2020origin}. Regardless of their origin, discovering additional Lempo-like systems will aid in understanding the formation of TNBs.


\subsubsection{Expectations for Non-Keplerian Effects}

While non-Keplerian fitting does have a variety of drawbacks and degeneracies (see further discussion in \citet{bpm1spinnymultimoon}), it is another powerful tool for understanding TNOs at a deeper level. Detecting non-Keplerian effects may allow for a new suite of unique measurements leading to a significant improvement in understanding TNBs, but the fraction of systems with detectable effects (with current or potentially future data) has been unknown. Non-Keplerian effects of multiple moons has been robustly detected in the Haumea system \citep[][]{2009AJ....137.4766R}, but no \jt{} precession has ever been robustly detected in any TNB system, despite attempts \citep[e.g.,][]{2009AJ....137.4766R,gourgeot2016near}. 

There is good reason, however, to expect that non-Keplerian shape effects are common and detectable. Firstly, aspherical shapes are common among TNOs, as shown by the results of occultation observations \citep[e.g.,][]{elliot2010size,benedetti2016results,ortiz2017size}. These shapes, assuming that TNOs are not all differentiated, should produce detectable non-Keplerian effects over the current observational baselines. Secondly, detection of \jt{} is common in both main-belt asteroids and near-Earth asteroids \citep[e.g.,][]{marchis2005discovery,marchis2005mass,fang2011orbits,fang2012orbits,vachier2012determination,beauvalet2014multiple,marchis2014puzzling}. For example, \citet{marchis2005mass} unambiguously detected the \jt{} of the asteroid (121) Hermione based on the orbit of its small companion, with observations spanning less than a single year. While TNBs generally have longer period orbits than asteroid binaries, several TNBs have relative astrometric measurements now reaching decade long timescales. The observational baselines and high quality data strongly suggest that non-Keplerian effects would be detectable in at least a few TNBs. 

To test the detectability of non-Keplerian effects in TNBs, we created synthetic relative astrometry for all TNBs, assuming the primary and secondary were interacting triaxial bodies with realistic shapes. The synthetic astrometry had realistic uncertainties added and was made to simulate the quantity and quality of real TNB relative astrometry by using existing datasets as templates. We then fitted the astrometry with Keplerian orbits to find if a Keplerian model provided adequate fits to the data. In many cases, we found that our synthetic astrometry was significantly inconsistent with a simple Keplerian model. We found the inconsistencies were strongest among TNBs with large primaries, although many near-equal sized binaries also exhibited inconsistencies. These results imply that non-Keplerian shape effects should be detectable and relatively common among TNBs. 

If it is indeed true that non-Keplerian effects are common in TNBs, it also raises questions about the validity of past results based on Keplerian orbital analyses. In most scenarios, Keplerian fitting should yield a similar orbit to a non-Keplerian fit, especially in the parameters not describing the orbit's orientation (e.g., total system mass, semi-major axis, and eccentricity). Some results, however, depend on the orientation of the orbits \citep[e.g.,][]{grundy2019mutual}, which slowly precess over time in a non-Keplerian framework. The systematic error introduced by this model misspecification needs to be evaluated to determine whether past research needs to be reevaluated. 

\subsection{Non-Keplerian Orbit Fitting}
\label{sec:orbitfitting}

For all of its benefits, orbit fitting of non-Keplerian orbits can be a difficult task. While the potential of triaxial body, given in Equation \ref{eqn:potential}, is relatively simple and could be integrated without much trouble\footnote{using Equations \ref{eqn:apsidal} and \ref{eqn:nodal} to calculate time-dependent orbital parameters is actually not much more efficient than full calculations of interacting quadrupoles \citep{bpm1spinnymultimoon}}, the actual dynamics of TNBs can be much more complicated. Not only can both a TNB's primary and secondary both have aspherical shapes with arbitrary rotation poles, but the gravitational interactions between the bodies cause a torque on those bodies, forcing precession of those rotation poles. The torques, which we call ``back torques,'' can make the dynamics of a TNB system much more complicated, especially when the binary is near-equal mass. Many past studies -- both observational and theoretical -- of non-Keplerian motion in asteroid binaries have neglected the effects of back torques. While this assumption is valid for very small secondaries (like many asteroid binaries), TNBs are often near-equal mass. The complexity involved in TNB spin-orbit dynamics requires the use of a coupled spin-orbit integrator, which can simultaneously integrate the translational and rotational equations of motion. 

For this reason, we developed a new spin-orbit integrator, named \spinny{}, which is able to self-consistently model coupled spin-orbit motion of an arbitrary number of gravitationally interacting triaxial ellipsoids, including the effect of back torques on the spinning bodies. \texttt{SPINNY} makes up the core of \multimoon, a package designed for non-Keplerian orbit fitting using state-of-the-art Bayesian techniques. Both \texttt{SPINNY} and \multimoon{} are described in full detail in \citet{bpm1spinnymultimoon} and are publicly available at \url{https://github.com/dragozzine/multimoon}, though we discuss the main ideas here. \multimoon's advanced statistical techniques and orbital model, combined with about a decade worth of relative astrometry for TNBs and extensive parallel computations, now make it possible to conduct a systematic survey to search for non-Keplerian shape effects in TNBs. 

Exploration of possible non-Keplerian parameters is generally a difficult task, requiring at least an order of magnitude more computational power than a simple Keplerian analysis. As such, the most interesting targets need to be identified to focus future work. Here, we perform a broad search for the best targets for more detailed non-Keplerian analyses. We focus our efforts on identifying TNBs with the most detectable and statistically significant non-Keplerian effects. 

\section{Methods}
In this work, for all orbit fitting procedures, we use \multimoon{}. \multimoon{} is a Python-based Bayesian orbit fitting package specifically designed to complete Keplerian and non-Keplerian orbit fitting of TNBs and other small solar system binaries \citep{bpm1spinnymultimoon}. For Keplerian orbit fitting, it uses the \texttt{SpiceyPy} \citep[][]{annex2020spiceypy}, a Python implementation of SPICE \citep[][]{acton1996ancillary}, to analytically solve the two-body problem. For non-Keplerian orbits, \multimoon{} uses \texttt{SPINNY} which numerically integrates the coupled spin and orbit equations of motion for an arbitrary number of interacting triaxial ellipsoids. These processes produce the position of the secondary relative to the primary. These are projected into the plane of the sky using ephemerides of the TNB system (relative to the Earth) from JPL Horizons, as queried by \texttt{astroquery} \citep[][]{ginsburg2019astroquery}. These ephemerides are corrected for light-time variations and other astrometric aberrations due to Earth's orbital velocity. Light-travel time and conversion to ecliptic coordinates is done as part of the data preparation. For more details, see \citet{bpm1spinnymultimoon} or the \multimoon{} code itself at \url{https://github.com/dragozzine/multimoon}. The analysis herein uses a version of \multimoon{} equivalent to Version 1.0. Since our fitting was completed, \multimoon{} has been updated to more accurately model the spin dynamics of small secondaries. We have confirmed that this small change has no impact on any of the results presented in this work. 

The Bayesian parameter inference at the heart of \multimoon{} is performed by \texttt{emcee} \citep[][]{foreman2013emcee,foreman2019emcee}, a Python-based Markov Chain Monte Carlo (MCMC) ensemble sampler. This allows us to robustly fit orbits in a Bayesian manner. \multimoon{} uses the same weighted least squares orbit fitting likelihood function that is ubiquitous in TNB orbit fitting \citep[e.g.][etc.]{grundy200842355, grundy2009mutual,2009AJ....137.4766R,kiss2019mass,holler2021eris}. \multimoon{} differs from past orbit fitters, however, in that it does not use a downhill optimization technique, but instead uses ensemble MCMC techniques that are robust to local minima. This avoids problems that have occurred when using problematic simple minimization techniques \citep{beauvalet2014multiple}. Data weights in our likelihood functions are equal to the inverse square of the measurement uncertainty.

Briefly, the outputs from \multimoon{} are a posterior probability distribution of inferred parameters given the observational data and priors (which are generally flat and uninformative). Based on the output posteriors, \multimoon{} produces a variety of different output plots, including both diagnostic plots (e.g., walker trace plots, likelihood plots, etc.) and publication-ready plots \citep[e.g., corner plots,][]{foreman2016corner}. For more discussion, see Ragozzine et al., submitted.

\subsection{Data}
\label{sec:data}
For all orbital fits completed in this paper, we use publicly available relative astrometric data and orbital fits catalogued in Will Grundy's online TNO database\footnote{\url{http://www2.lowell.edu/users/grundy/tnbs/status.html}, data retrieved September 26, 2020}, providing the relative astrometry and orbital solutions of 45 TNBs with solved or mirror ambiguous orbits. The list of all the TNBs in our analysis is given in Table \ref{tab:tnbs}; note that the meaning of some columns is explained in more detail below. With both relative astrometry and orbital solutions in hand, both are converted into ecliptic coordinates and used in our analysis as described above. All inputs and outputs from our fits are available online on Zenodo (\dataset[doi: 10.5281/zenodo.10620251]{https://doi.org/10.5281/zenodo.10620251}{}).

Although the majority of our sample consists of the TNBs whose mutual orbits have been previously studied in the literature, we exclude two systems, (136199) Eris-Dysnomia and (469514) 2003 QA$_{91}$. The Eris-Dysnomia system, whose Keplerian orbit has been extensively studied in the past \citep[][]{2007Sci...316.1585B,holler2021eris}, is known to be $>6\sigma$ inconsistent with a single Keplerian orbit \citep[][]{holler2021eris}. Non-Keplerian orbit fitting (completed with \multimoon{}) is discussed in Spencer et al., in prep. For 2003 QA$_{91}$, the publicly available relative astrometry \citep[which are compiled in][]{grundy2019mutual} are clearly inconsistent with the given orbit solution. We believe that a typographical error is present in the publicly available astrometry. For these reasons, we exclude these systems from our analysis. 

Most of the relative astrometry we use is derived from precise images from HST, although significant contributions are made from a variety of other ground-based observatories. Typical error bars on the astrometry from HST are less than 10 milliarcseconds, although they can even be sub-milliarcsecond when observations were taken with HST's High Resolution Camera. We also use many observations from Keck using its adaptive optics system, which typically give measurements with similar precision to HST. For some of the wider TNBs, observations can be obtained with smaller ground-based telescopes like Gemini, Magellan, and the Canada France Hawaii telescope. Given the lower resolution of these telescopes, typical uncertainties can be 100s of milliarcseconds. To get a sense of the typical precision of the observations used for our orbit fitting, we list the median uncertainty of the astrometric measurements in Table \ref{tab:tnbs}.

For all astrometric data, we assume that uncertainties are Gaussian and measurements are independent. Unfortunately, when combining various data sources (i.e., different telescopes and instruments), systematic errors can arise from cross-calibration issues, uncertainties on time varying plate solutions, and other issues. Past analyses have shown that these issues are not significant for the vast majority of TNOs \citep[e.g.,][etc.]{grundy200842355,grundy2014orbit,grundy2019mutual}, most of which are not resolved (or barely resolved) in even the largest of telescopes. Further, \textit{a posteriori}, we can use heuristics like reduced $\chi^2$ to check whether our noise model (and orbit model) adequately describes the data. We find that systematics in our data are unlikely to affect our modeling significantly and add noise at (or below) the measurement uncertainties.

\begin{deluxetable*}{lCcCcCC}
\centerwidetable
\tabletypesize{\footnotesize}
\tablecaption{System Properties of TNBs}
\tablehead{
Object & \text{Primary Rotation} & Reference & \text{Radius} & Reference & \text{Epoch} & \text{Uncertainty}\\
 & \text{Period (h)} & & \text{(km)} & & \text{(JD)} & \text{(arcsec)}
}
\startdata
(26308) 1998 SM$_{165}$ & 8.4 & \cite{spencer2006low} & 134 & \cite{stansberry2008physical} & 2452700.0 & 0.001 \\
(42355) Typhon-Echidna & 9.67 & \cite{duffard2009transneptunian} & 81 & \cite{santos2012tnos} & 2454000.0 & 0.003 \\
(50000) Quaoar-Weywot & 8.839 & \cite{fornasier2013tnos} & 535 & \cite{fornasier2013tnos} & 2454000.0 & 0.008 \\
(55637) 2002 UX$_{25}$ & 14.382 & \cite{rousselot2005photometric} & 332.5 & \cite{fornasier2013tnos} & 2454000.0 & 0.002\\
(58534) Logos-Zoe & \nodata & \nodata & 41 & \cite{grundy2011five} & 2452600.0 & 0.003 \\
(60458) 2000 CM$_{114}$ & \nodata & \nodata & 83.5 & assumed & 2457000.0 & 0.004 \\
(65489) Ceto-Phorcus & 4.43 & \cite{dotto2008rotational} & 111.5 & \cite{santos2012tnos} & 2453880.0 & 0.003 \\
(66652) Borasisi-Pabu & 6.4 & \cite{kern2006study} & 63 & \cite{vilenius2014tnos} & 2451900.0 & 0.002 \\
(79360) Sila-Nunam & 300.2388 & \cite{thirouin2014rotational} & 124.5 & \cite{vilenius2014tnos} & 2454400.0 & 0.004 \\
(80806) 2000 CM$_{105}$ & \nodata & \nodata & 80 & assumed & 2457000.0 & 0.004 \\
(88611) Teharonhiawako-Sawiskera & 9.505 & \cite{osip2003physical} & 89 & \cite{vilenius2014tnos} & 2452000.0 & 0.015 \\
(90482) Orcus-Vanth & 13.188 & \cite{rabinowitz2006diverse} & 458.5 & \cite{brown2018medium} & 2454000.0 & 0.001 \\
(119979) 2002 WC$_{19}$ & \nodata & \nodata & 169 & \cite{lellouch2013tnos} & 2457000.0 & 0.006 \\
(120347) Salacia-Actaea & 6.5 & \cite{thirouin2014rotational} & 427 & \cite{brown2017density} & 2454300.0 & 0.005 \\
(123509) 2000 WK$_{183}$ & \nodata & \nodata & 58 & \cite{grundy2011five} & 2454300.0 & 0.003 \\
(134860) 2000 OJ$_{67}$ & \nodata & \nodata & 69 & \cite{grundy2009mutual} & 2457000.0 & 0.003 \\
(148780) Altjira & \nodata & \nodata & 123 & \cite{vilenius2014tnos} & 2454300.0 & 0.002\\
(160091) 2000 OL$_{67}$ & \nodata & \nodata & 37 & assumed & 2457000.0 & 0.011 \\
(160256) 2002 PD$_{149}$ & \nodata & \nodata & 93 & assumed & 2457000.0 & 0.007 \\
(174567) Varda-Ilmarë & 5.91 & \cite{thirouin2014rotational} & 370 & \cite{vilenius2014tnos} & 2455300.0 & 0.005 \\
(225088) Gonggong-Xiangliu & 44.81 & \cite{pal2016large} & 767.5 & \cite{pal2016large} & 2457000.0 & 0.006 \\
(229762) G!kún\textdoublepipe’hòmdímà-G!ò’é !hú & 11.05 & \cite{thirouin2014rotational} & 295 & \cite{schindler2017results} & 2457000.0 & 0.004 \\
(275809) 2001 QY$_{297}$ & 11.68 & \cite{thirouin2012short} & 84.5 & \cite{vilenius2014tnos} & 2454200.0 & 0.002 \\
(341520) Mors-Somnus & 9.28 & \cite{thirouin2014rotational} & 51 & \cite{sheppard20122007} & 2454300.0 & 0.030 \\
(364171) 2006 JZ$_{81}$ & \nodata & \nodata & 61 & \cite{parker2011characterization} & 2453400.0 & 0.026 \\
(385446) Manwë-Thorondor & 11.88 & \cite{rabinowitz2019complex} & 80 & \cite{grundy2014orbit} & 2454400.0 & 0.004 \\
(469705) \kagara{} & 9.65 & \cite{benecchi2013light} & 69 & \cite{vilenius2012tnos} & 2455600.0 & 0.005 \\
(508788) 2000 CQ$_{114}$ & \nodata & \nodata & 66 & assumed & 2454100.0 & 0.003 \\
(508869) 2002 VT$_{130}$ & \nodata & \nodata & 125.5 & \cite{mommert2013remnant} & 2455900.0 & 0.002 \\
(524366) 2001 XR$_{254}$ & \nodata & \nodata & 85.5 & \cite{vilenius2014tnos} & 2454300.0 & 0.002 \\
(524531) 2002 XH$_{91}$ & \nodata & \nodata & 149 & assumed & 2457000.0 & 0.005 \\
(525462) 2005 EO$_{304}$ & \nodata & \nodata & 152 & \cite{parker2011characterization} & 2453400.0 & 0.042\\
(612095) 1999 OJ$_{4}$ & \nodata & \nodata & 37.5 & \cite{grundy2009mutual} & 2454000.0 & 0.003 \\
(612147) 2000 CF$_{105}$ & \nodata & \nodata & 32 & \cite{parker2011characterization} & 2454881.0 & 0.024 \\
(612176) 2000 QL$_{251}$ & \nodata & \nodata & 74 & \cite{grundy2009mutual} & 2454200.0 & 0.003 \\
(612239) 2001 QC$_{298}$ & 7.78 & \cite{thirouin2014rotational} & 117.5 & \cite{vilenius2014tnos} & 2452000.0 & 0.002\\
(612687) 2003 UN$_{284}$ & \nodata & \nodata & 62 & \cite{parker2011characterization} & 2452963.8 & 0.072 \\
1998 WW$_{31}$ & \nodata & \nodata & 74 & \cite{2002Natur.416..711V} & 2452300.5 & 0.006\\
1999 RT$_{214}$ & \nodata & \nodata & 50 & assumed & 2457000.0 & 0.005 \\
2001 QW$_{322}$ & \nodata & \nodata & 64 & \cite{parker2011characterization} & 2452000.0 & 0.076 \\
2003 QY$_{90}$ & 3.4 & \cite{kern2006study} & 40.5 & \cite{grundy2011five} & 2453500.0 & 0.006 \\
2003 TJ$_{58}$ & \nodata & \nodata & 32.5 & \cite{grundy2009mutual} & 2454300.0 & 0.003 \\
2004 PB$_{108}$ & \nodata & \nodata & 121.5 & \cite{grundy2011five} & 2454200.0 & 0.003 \\
2006 BR$_{284}$ & \nodata & \nodata & 45 & \cite{parker2011characterization} & 2455000.0 & 0.031 \\
2006 CH$_{69}$ & \nodata & \nodata & 50 & \cite{parker2011characterization} & 2455000.0 & 0.034 \\ 
\hline
\enddata
\tablecomments{The list of 45 Trans-Neptunian Binaries which we study with both Keplerian and non-Keplerian models. This includes all TNBs (except two unusual cases) that had known orbits or mirror-ambiguous orbits. Rotation periods and primary radii used in our non-Keplerian fits (and their references) are also listed. The epoch of the fit is also listed; this is the time at which the orbital element parameters are set in non-Keplerian fits (which have time-varying orbital elements). Uncertainty refers to the typical uncertainty in the relative astrometry used to produce orbit fits. }
\label{tab:tnbs}
\end{deluxetable*}

\subsection{Keplerian Orbital Fits}
\label{sec:kep_methods}

As a preliminary step to completing non-Keplerian fits, we first completed a round of Keplerian fits to validate our orbit fitting techniques and provide full Bayesian posteriors for the Keplerian orbits of TNBs. This allows us to analyze the quality of the Keplerian fits and provide a baseline for comparison when completing non-Keplerian fits. 

Our Keplerian orbital model has seven parameters, including the usual six orbital elements ($a$, $e$, $i$, $\omega$, $\Omega$, and $\mathcal{M}$) and total system mass ($M_{sys}$). All angles are referenced to the J2000 ecliptic plane (unlike many publications of these TNB orbits which use angles in the equatorial reference frame). We also consider a model with constant photocenter-barycenter offests which is discussed in Appendix \ref{sec:kepoff_methods}. 

Our Keplerian fits were run with 100 walkers in the MCMC ensemble, with a 5000 step burn in, pruning of walkers significantly far away from the best parameter space \citep[similar to][]{proudfoot2019modeling,hou2012affine}, a 1000 step post-pruning burn in, and a 5000 step sample. We initialized walker positions for all parameters by drawing random, normally-distributed samples from the orbital solutions listed in the Grundy database, although we inflated uncertainties to allow for a broader search of parameter space. For systems with mirror ambiguous orbits, we ran two different fits to explore both orbital solutions.

We set uninformative priors for these fits, using uniform distributions for each parameter (see Ragozzine et al. for more details). Similar to other orbit fitting procedures, we compared our model to the relative astrometry using the $\chi^2$ statistic \citep[c.f.][]{grundy2019mutual}. After completion of the runs, we checked for convergence of the MCMC chains based on trace plots and the smoothness of the marginal and joint posterior distributions of each parameter. In a few cases where we could not clearly confirm the MCMC chains were converged, we reran the fits with more burn in and sampling steps, until we were confident the chains were converged.

\subsection{Non-Keplerian Orbital Fits}
\label{sec:nonkep_methods}

Using \multimoon's non-Keplerian orbit fitter, we completed non-Keplerian fits to all 45 TNB systems. Although \multimoon{} is capable of fitting the gravitational harmonics of all bodies in TNBs system, for this analysis, we only fit the \jt{} gravitational harmonic of the system's primary with an assumed size and rotation rate. This greatly reduces the complexity of the model as only 3 parameters are added (\jt{} and two spin pole direction angles) rather than 14 (ellipsoid polar axis length $c$, \jt, \ct, two spin pole direction angles, a spin longitude angle, and a rotation rate, all for both objects). Since non-Keplerian effects can break the usual mass degeneracy in Keplerian fits, we also allow both masses to float, adding one more parameter. In many TNB systems, the dominant source of non-Keplerian effects should come from the \jt{} of the primary, but even if this is not the case, using a single \jt{} is a good approximation to the modeling the system's ``total'' \jt, as measured by the orbital precession. It also provides a way to model non-Keplerian orbital precession whatever its source, as discussed above and in \citet{bpm1spinnymultimoon}. While neglecting \ct{} can provide worse fits, this is only the case when a TNB is near a low-order spin orbit resonance. Given that most TNOs rotate in $\sim$10s of hours and TNB orbit periods are $\gtrsim$10 days, low-order spin-orbit resonances among our sample will be extremely rare. The only known TNB in our sample which is at spin-orbit resonance is Sila-Nunam. We discuss the consequence of neglecting \ct{} for the Sila-Nunam system in Section \ref{sec:nonkeplerian_fits}. As our goal is identifying objects with the most statistically significant non-Keplerian effects, even if our assumptions do not hold, the improvement in orbital fit from the wrong model would still indicate the need for higher fidelity and more advanced fits. 

Given these assumptions, our non-Keplerian orbital model has 11 free parameters, the six Keplerian orbital elements at a given epoch, the masses of the individual system components ($M_1$ and $M_2$), two angles describing the rotation pole of the primary at epoch ($i_{sp}$ and $\Omega_{sp}$, which are Euler angle representations of the spin pole), and $\ln\left(J_2R^2\right)$. We opt to use the $\ln$ of $J_2R^2$ to enable easier exploration over many orders of magnitude. Additionally, we use the combination $J_2R^2$ instead of just \jt{} so results can be interpreted with a variety of shape models. 

In our MCMC ensembles, we used 100 walkers running for 26000 total steps, split between 15000 burn in steps, pruning of poorly performing walkers, 1000 post-pruning burn in steps, and 10000 sampling steps. We set integration tolerance value at $10^{-10}$ for all runs. This was chosen to balance integration quality/accuracy and computational expense. In all our testing, this tolerance level was found to be sufficient for our needs. 

We initialized walker positions for mass and orbital elements identically to the Keplerian fits. In addition, we initialized walker positions for $\log\left(J_2R^2\right)$ by drawing random samples from a normal distribution centered at $\log\left(10R\right)$, where $R$ is the object's estimated radius, a rough approximation of typical \jt values due to non-spherical shapes over a large size range. Lastly, we initialized walker positions for the spin pole direction to be generally aligned with the orbit, within $\sim$10-30$\degr$ of perfectly aligned. For other required \multimoon{} inputs (approximate radius and rotation period), we used values previously published in the literature (see Section \ref{sec:data} for more details). For objects with no known rotation period, we used a default value of 10 hours \citep[similar to those found in ][]{thirouin2014rotational}. For the few TNBs that did not have estimated radii published in the literature, we assumed a value based on an assumed albedo and the system's absolute magnitude. Information on all TNBs considered in our analysis, along with the input for approximate radius and rotation period, is located in Table \ref{tab:tnbs}. Our testing indicates that changing these assumed values does not substantially affect the results of our fits. Like the Keplerian orbit fits, mirror ambiguous TNBs had two orbit fits run to explore both orbital solutions. 

Our priors on each parameter were uniform with the goal of being uninformative, except in a few cases. Firstly, we enforced $a(1-e) > q_{min}$, where $q_{min}$ was chosen to be equal to a factor of a few times the primary radii. This helped to reduce the chances of unphysical close encounters that were occasionally explored in extreme non-Keplerian integrations. 
We also force $M_2 < M_1$. This practice reduces the effects of degeneracies between mass and \jt. 

Finally, $\ln\left(J_2R^2\right) < 15$ was enforced to limit exploration in $J_2$ space. At large values of $J_2R^2$, some TNB systems can become unstable, placing this constraint prevents exploration of these unstable models. While this may prevent exploration of relevant parameter space for some systems, we found that this prior appropriately balances stability and our goal of broad exploration. In a small number cases, the prior had to be further reduced after confirming that large values of \jt{} caused unphysical models\footnote{These unphysical models could sometimes cause \multimoon{} to crash unexpectedly, this has since been fixed.}. A uniform prior in  $\ln\left(J_2R^2\right)$ is not meant to signify our actual prior knowledge of the \jt{} distribution of TNBs, but rather to encourage exploration of this new parameter. We adopt as meaningful only those fits where the likelihood (calculated based on $\chi^2$ as with Keplerian modeling) of a non-Keplerian fit strongly prefers a particular value of \jt and thus is relatively independent of the prior. 

As our goal of this project was to identify systems with the most statistically significant non-Keplerian effects, we did not aim for full convergence of our MCMC chains. Rather, we focused on achieving the best non-Keplerian fit possible in a fixed number of model evaluations. While this results in unconverged fits -- so that the full posterior probability distributions are not expected to be completely accurate -- the unconverged chains can still be informative, especially in light of our goal of identifying targets for future investigations. In essence, our runs are set up to determine which TNBs have the most easily detectable non-Keplerian effects. We refer to these potentially unconverged fits as ``exploratory'' and consider them valuable at the full catalog level of this analysis. 

These fits and their implications were studied individually in detail by various co-authors as part of Brigham Young University's Physics 227 (``Solar System Astronomy'') Class Project. 

\subsection{Non-Keplerian Orbital Fits for Borasisi-Pabu}

To show that unconverged fits described above are able to identify targets worthy for future investigation, we completed a full non-Keplerian orbit analysis of the Borasisi-Pabu system, one of the more promising targets identified in our exploratory fits, as a proof-of-concept. In this fit, we relaxed several of the assumptions made in the non-Keplerian exploratory fits, most notably our assumptions regarding the \ct{} of the primary. Our non-Keplerian orbital model has 13 parameters, six Keplerian orbital elements at epoch, the masses of both system components, the direction of the primary's rotation pole at epoch, the $J_2R^2$ and $C_{22}R^2$ of the primary, and the longitude of the primary's prolate axis at epoch ($\omega_{sp}$). For Borasisi's rotation period we used 6.4 hours \citep[][]{kern2006study} and a radius of 63 km \citep[][]{vilenius2014tnos}. After our analysis was completed, \citet{kecskemethy2023light} found a different light curve period for the combined system, we discuss the minor implications of this in Section \ref{sec:borasisi}.

For this full orbit fit, we ran \multimoon{} with 980 walkers for 23500 steps, split between 15000 burn in steps, 1000 post-pruning burn in steps, and 7500 sampling steps. Integration tolerance was set to $10^{-11}$ to produce the best quality of fits possible, with less regard for optimizing computational expense than the previous round of fits. Initial walker positions were drawn based on preliminary orbital fits and the results of our exploratory fits, but otherwise the fitting process was effectively identical to that above. We confirmed convergence of the MCMC chain by inspection of trace plots and marginal and joint posterior distributions of each parameter.


Our quadrupolar approximation cannot distinguish between models rotated by 180 degrees (e.g., it cannot distinguish the North pole from the South pole), resulting in a two-fold degeneracy. Since the degeneracy is fairly well-understood, we decide to fit only the prograde rotation solution, where the rotation axis is required to be inclined $<90\degr$ relative to the binary's mutual orbit plane. 


\subsection{Evaluating Fit Quality}
To ensure the quality of all of our fits (both Keplerian and non-Keplerian), we have thoroughly evaluated each fit using a variety of statistical techniques. Most simply, and easily output from \multimoon{} is the calculation of reduced $\chi^2$, which is equivalent to the best fit parameter set's $\chi^2$ per degree of freedom. For almost all of our fits, reduced $\chi^2$ was $\lesssim1$, indicating good quality fits. For 8 of our Keplerian fits, however, we found statistically significant cases of elevated reduced $\chi^2$. Several of these systems have been previously identified as inconsistent with Keplerian orbits \citep[e.g.][]{grundy2019mutual}. These systems all had significant, or nearly significant, non-Keplerian effects detected (see Section \ref{sec:nonkeplerian_fits}). 

Another way we can analyze the quality of our orbit fits is to calculate the root-mean-square (RMS) residual of the fit. For our fits, we find that most of our fits have RMS residuals $\lesssim 15$ milliarcseonds, indicating an excellent fit to the data. Compared with the typical uncertainties in the astrometry we fit to (see Table \ref{tab:tnbs}), we find that our fits are very robust. A few systems have much larger RMS residuals, but this is caused by large uncertainties in the data. As discussed in Section \ref{sec:data}, the ultra-wide TNBs tend to have many observations taken with relatively imprecise ground-based observatories, which can result in large RMS residuals. Combined RMS residuals (with each dimension added in quadrature) are reported in all tables below.

Like RMS residuals, we also looked at the weighted RMS residuals. These residuals are essentially in units of observational error bars (as shown in Table \ref{tab:tnbs}). We find that all of our fits have weighted RMS residuals $<1.4$, with most being $<1$. This confirms the quality of our fits. 

We also closely examined the residual plots for each fit (see Figures \ref{fig:kep_residual} and \ref{fig:nonkep_residual} for examples). These plots show the normalized residuals (residual divided by measurement uncertainty) and were examined to ensure no systematic trends were visible in the residuals (e.g., uncentered residuals, significantly larger residuals in one dimension, correlations among subsets of the data taken with different facilities, etc.). We found no serious issues with any of our fits in individual fits, and also found no problems at the ensemble level. These normalized residuals are available for both our Keplerian and non-Keplerian fits publicly available on Zenodo \trackchange{(\dataset[DOI: 10.5281/zenodo.10620251]{https://doi.org/10.5281/zenodo.10620251})}.

Another sign of the quality of our fits is their close match to orbit fits published in the literature. In Section \ref{sec:keplerianfits}, we show that our orbit fits closely reproduce fits in the literature. This close agreement is strong evidence that our orbit fits are of high quality.

\begin{figure*}
\centering
\includegraphics[width=\textwidth]{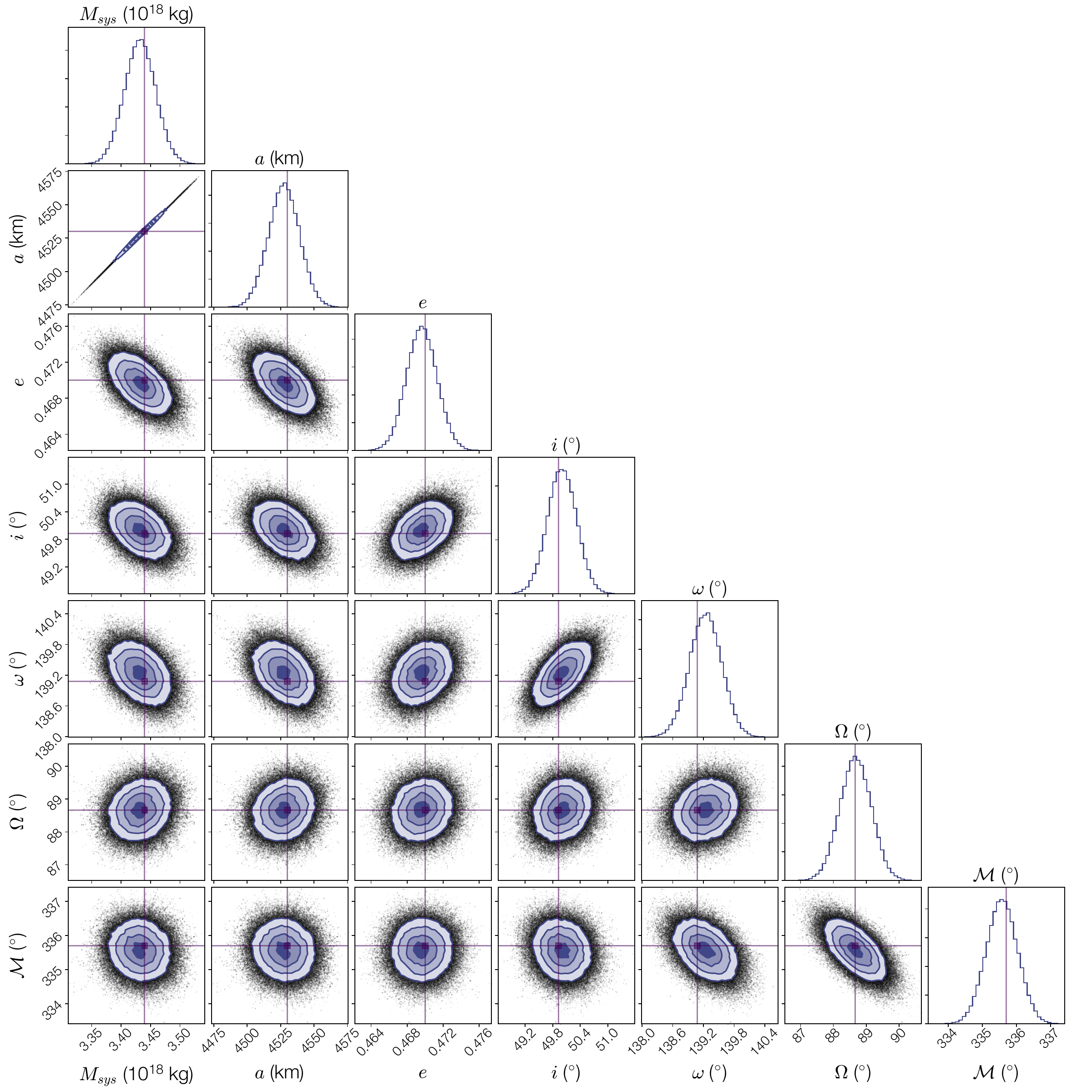}
\caption{Corner plot for the Borasisi-Pabu Keplerian orbit fit. Along the tops of the columns are the marginal posterior distribution for each parameter; in the case of a Gaussian distribution, this can be used to determine the best-fit and uncertainty. The contour plots show the joint posterior distribution between every pair of parameters. For example, the system-mass semi-major axis degeneracy is seen in the top left which is a result of measurement of only the orbital period and Newton's Version of Kepler's Third Law. Contours show the 1, 2, and 3$\sigma$ levels of each joint posterior distribution. The horizontal and vertical lines show the orbit solution publicly available in the Grundy database. Of particular interest is the strong agreement between the Grundy solution (horizontal and vertical lines) and the solution derived by \multimoon. The best fit in the \multimoon{} fit corresponded to a $\chi^2$ of $\sim$21. RMS residuals are 2 milliarcseconds in both longitude and latitude. All angles are relative to the J2000 ecliptic plane on JD 2451900.0 (2000 December 21 12:00 UT).}
\label{fig:borasisi-kep}
\end{figure*}

\begin{figure}
\centering
\includegraphics[width=\linewidth]{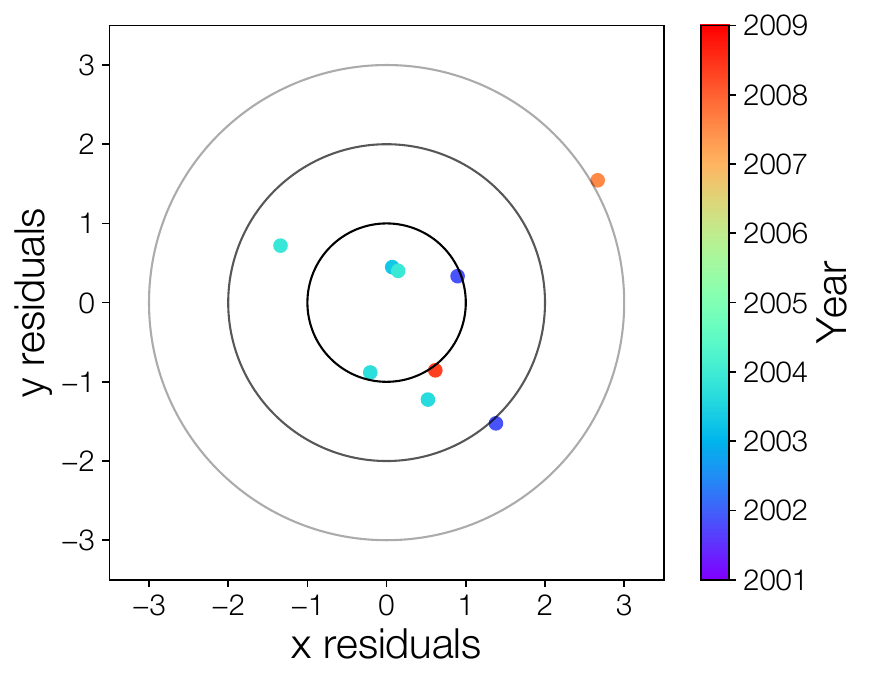}
\caption{Residual plot for the Borasisi-Pabu Keplerian orbit fit. Plotted are the normalized residuals (residual divided by measurement uncertainty) in longitude ($x$) and latitude ($y$) with color corresponding to observation date. The three large circles correspond to 1, 2, and 3 $\sigma$ error contours. The best fit in the \multimoon{} fit corresponded to a $\chi^2$ of $\sim$21. RMS residuals are 2 milliarcseconds in both longitude and latitude. The radius of a typical 1$\sigma$ error contour is $\sim$2 milliarcseconds, although values ranged between 0.9 and 5 milliarcseconds. }
\label{fig:kep_residual}
\end{figure}


\section{Keplerian Fits}
\label{sec:keplerianfits}

The results of our Keplerian fits are contained in Table \ref{tab:keplerian}. It contains information regarding the posterior distributions of the mass and orbital elements of each TNB (median values with $1\sigma$ confidence intervals). In addition to values from the marginal (single variable) posterior distributions (as seen in the table), the full posterior distribution can be displayed as a corner plot. As an example, we show the corner plot of the Keplerian posterior for (66652) Borasisi-Pabu in Figure \ref{fig:borasisi-kep} and a residual plot in Fgiure \ref{fig:kep_residual}. The results for all 45 TNBs (including MCMC chains, diagnostic plots, statistical information, etc.) are publicly available\trackchange{\footnote{\dataset[DOI: 10.5281/zenodo.10620251]{https://doi.org/10.5281/zenodo.10620251}}}. 

For all mirror ambiguous TNBs, two orbit solutions are shown in Table \ref{tab:keplerian}, with the exception of 2002 WK$_{183}$ and 2000 OJ$_{67}$. For both of these objects, despite attempting two orbit fits for the separate solutions, the mirror ambiguous solutions are so close together that the posteriors of each solution significantly overlap. Separating the two distributions in a statistically rigorous way is difficult due to the blended nature of the posterior. Note that the \textit{emcee} algorithm at the heart of \multimoon{} is not optimized for rigorous exploration of multi-modal posterior distributions. As such, we report the statistics of the blended distribution and urge caution when using the orbit fits for these objects.

\begin{figure}
\includegraphics[width=\linewidth]{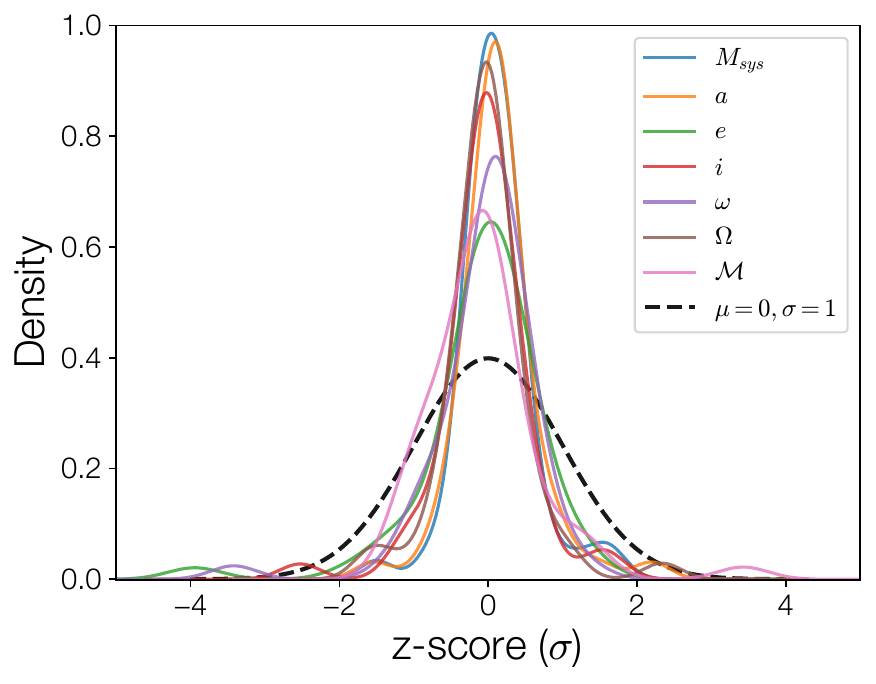}
\caption{Kernel density estimates (KDEs) for the distributions of \textit{z}-scores of Grundy best fits (when compared to the \multimoon{} posteriors) for each fitted parameter. The dashed black line shows a normal distribution with a width of 1 as a comparison. It is clear that for all fitted parameters, the Grundy best fits are well within our posterior distributions, with the majority being $<1\sigma$ away from our posteriors. \multimoon{}'s Keplerian fitting algorithm produces essentially identical results to those previously published in the literature.}
\label{fig:grundycomp}
\end{figure}

As a test of \multimoon's performance in completing Keplerian fits, the posterior distributions of the ensemble of fits can be compared to the best fits from the Grundy database. We calculated the \textit{z}-score of every Grundy best fit for every parameter and every object, using our posterior distribution as the reference distribution. The distributions of these scores for each parameter is shown in Figure \ref{fig:grundycomp} using kernel density estimates. The \textit{z}-score distributions clearly show that the Grundy best fits are consistent with the \multimoon{} posteriors. Technically, since we are analyzing the same data with the same model, the agreement should be much better than drawing parameters from a random distribution. While not a perfect match, the fact that the vast majority of parameters are within 1-$\sigma$ means that any systematic errors are minimally important compared to the statistical errors. We note that a similar analysis of an ensemble of Keplerian TNB orbit fits also found consistency with the Grundy results \citep{2020MNRAS.494.2410E}, further supporting that these results are robust to the analysis method. 

Likewise, when comparing the Grundy database parameter uncertainties (which are determined via Monte Carlo techniques) to the \multimoon{} Keplerian posterior distributions, we find excellent agreement. For all parameters (except argument of periapsis $\omega$ and mean anomaly $\mathcal{M}$), the error bars are similarly sized for all the TNBs. For $\omega$ and $\mathcal{M}$, our model is differently parameterized, where the Grundy database uses longitude of periapse ($\varpi$) and longitude at epoch ($\epsilon$). This slightly different parameterization obscures a like-to-like comparison. 

For those interested in precise understanding of the uncertainties in Keplerian fit parameters, our reporting of full posteriors provides an improvement to the state-of-the-art which approximates parameters as having a mean value and uncertainty. Posterior samples are also particularly useful for scheduling of follow-up observations. Each sample in the chain provides a predicted position of the binary components at any given time. Taken together, the ensemble of samples (or a sufficiently large subset of the ensemble) then provides a probability distribution for the predicted positions of the binary components, as further discussed below. 

The results for the Keplerian model with constant photocenter-barycenter offsets are discussed in Appendix \ref{sec:kepoff_methods}. We find that the offsets presented are attributable to overfitting. We also find that they do not significantly affect the Keplerian results, except for possibly adjusting the eccentricity. 

The agreement of our Keplerian outputs with previously published fits confirms that \multimoon{} can effectively fit TNB orbits, despite the different parameterization and methods used. This validates \multimoon's fitting procedures and techniques, lending confidence in our non-Keplerian fits presented in Sections \ref{sec:nonkeplerian_fits} and \ref{sec:borasisi}.

Although our Keplerian fits only reproduce past work, this work allows us to identify possible systematic issues with TNB orbit fitting at the catalogue level. Furthermore, it allows us to understand if the uncertainties reported in astrometric data analysis are accurate. Given our high-quality fits, we find that Keplerian orbit fitting and current generation astrometric data analysis adequately account for systematic effects. We measure this using fit quality heuristics like reduced $\chi^2$ and p-values. Aside from a few systems that are expected to have non-Keplerian orbits (e.g., Salacia, Orcus, etc.), our fits give reduced $\chi^2 \lesssim1$, which implies that the reported astrometric uncertainties and orbit model describe the data well. Although this does not preclude all systematic effects, it significantly reduces the number of possible sources of error. Future investigations into sources of systematic error should use flexible noise models that can parameterize the sources of error present in the astrometric data. 


\section{Non-Keplerian Fits}
\label{sec:nonkeplerian_fits}

Aligned with our goal of identifying candidates for future non-Keplerian analysis, we present the non-Keplerian best fit model for each TNB in Table \ref{tab:nonkeplerian}. The best fits in the table do not have uncertainties attached to them because our MCMC chains are not converged. In an MCMC framework, uncertainties are drawn from the 16th and 84th percentiles of the output chains. Without convergence, the resulting 16th and 84th percentiles are extremely unreliable estimations of the true uncertainties. Despite this, we still believe that the best fits are still useful for showing that non-Keplerian effects are detectable. To evaluate the detectability of non-Keplerian effects, we use a likelihood ratio test comparing the Keplerian and non-Keplerian orbit fits. The likelihood ratio test compares the goodness-of-fit of two nested models and quantifies the improvement in fit quality between the models. In our case, the Keplerian model is a subset of a non-Keplerian model where $J_2 = 0$. Under the likelihood ratio test if the null hypothesis (no currently detectable non-Keplerian effects) is supported by the data, the likelihood ratio, $\frac{\mathcal{L}_K}{\mathcal{L}_{NK}} \approx 1$, where ${\mathcal{L}_K}$ and ${\mathcal{L}_{NK}}$ are the Keplerian and non-Keplerian likelihoods, respectively. If the null hypothesis can be rejected (non-Keplerian effects are currently detectable), $\frac{\mathcal{L}_K}{\mathcal{L}_{NK}} \ll 1$.

Traditionally, some threshold is chosen at which the improvement in fit is deemed great enough and the null hypothesis is formally rejected. However, the fits presented here are exploratory in nature and may not have reached the global maximum likelihood, as evidenced by our results in Section \ref{sec:borasisi}. Further analysis would only increase the improvement of non-Keplerian models, as these results act as a lower limit on the improvement of the fit. In keeping with our goal to identify targets for future analysis, we choose to give special attention to cases where $\frac{\mathcal{L}_K}{\mathcal{L}_{NK}} < 0.1$ which are bolded in Table \ref{tab:nonkeplerian}. Inspection of the posterior distributions (though not necessarily converged) also show that in these systems, $J_2R^2 = 0$ is disfavored. Further work (i.e. converged fits) is needed to truly determine the true statistical significance of these improvements, although we confirm the improvement in the likelihood is more than expected from the addition of new parameters. 

We do find that change in RMS residuals between our Keplerian and non-Keplerian fits are not strongly correlated with improvement in fit as measured by $\frac{\mathcal{L}_K}{\mathcal{L}_{NK}}$. Some exploration of possible reasons as to why this is the case showed that the result is due to heterogeneous uncertainties in the data. When instead using normalized/weighted residuals (as in Figure \ref{fig:kep_residual}), we see the expected correlation between change in fit quality and change in weighted RMS residuals.

The 8 TNBs satisfying our improvement threshold have non-Keplerian effects that are currently detectable. In this section, we will first discuss population-level trends identified in our fits, and then individually discuss the 8 TNBs we identify as prime targets for non-Keplerian analysis. 

\begin{figure*}
    \centering
    \includegraphics[]{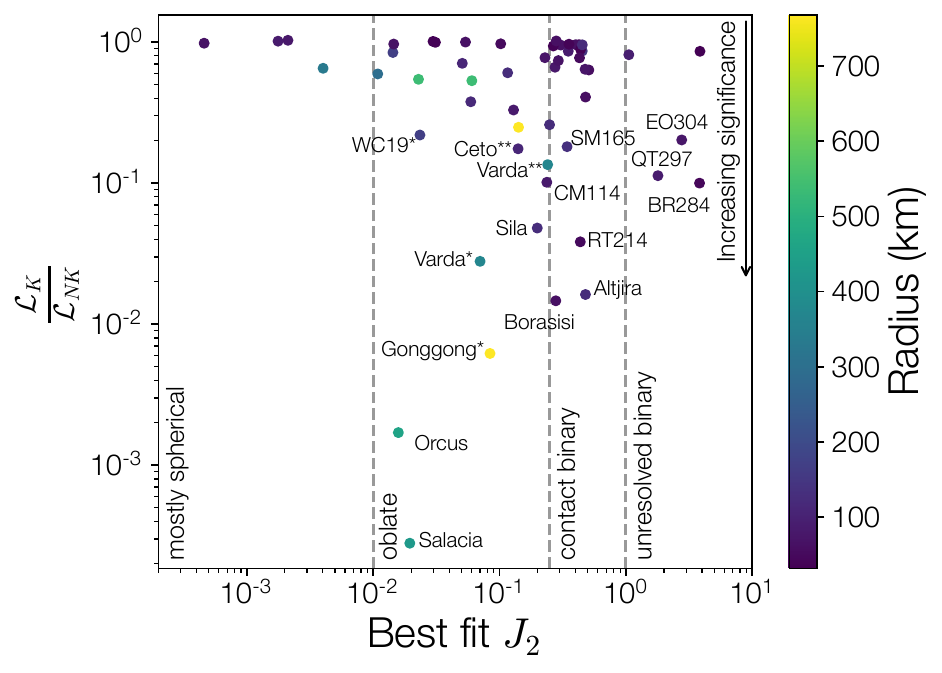}
    \caption{Improvements in fit quality by adopting a non-Keplerian model. The best fit \jt{} presented here is found by assuming a triaxial shape and dividing our measured $J_2R^2$ by the resulting body's volumetric radius squared. The ratio $\frac{\mathcal{L}_K}{\mathcal{L}_{NK}}$ compares the Keplerian models presented in Section \ref{sec:keplerianfits} to the best exploratory fit presented in this section. If the likelihood ratio is $\ll1$, the null hypothesis (no detectable non-Keplerian effects) could be rejected, and non-Keplerian effects are indeed detected. In general, throughout this paper, we adopt a threshold of $\frac{\mathcal{L}_K}{\mathcal{L}_{NK}} < 0.1$ to indicate systems which have orbit fit improvements indicative of non-Keplerian motion. We have labeled all of our systems with improvements below our threshold, as well as several systems with marginal improvements. Asterisks in labels indicate either a prograde (*) or retrograde (**) orbital solution for mirror ambiguous TNBs. Color corresponds to the radius of the TNB's primary. We find that many TNBs are far better explained by a non-Keplerian model, with many systems showing moderate improvements. We also find a wide range of acceptable \jt{} values, ranging from oblate spheroids to unresolved binaries. Future work and observations should focus on confirming the significance of our detections and extending the observational baselines of systems with moderate improvements in fit quality. We urge caution in using the values of \jt{} in this plot as there may be significant uncertainties that our exploratory fits are unable to fully constrain/measure. }
    \label{fig:j2-sig}
\end{figure*}

In addition to Table \ref{tab:nonkeplerian}, we display the ensemble of non-Keplerian fits in Figure \ref{fig:j2-sig} where we plot the best fit \jt{} value against the likelihood-ratio statistic. In this figure, we assume a triaxial shape model for easy comparison between our fits and to the literature. We emphasize that only the best fit orbit (or \jt) is shown, as our fits are not completely converged. As such, no uncertainties can be associated with these measurements. Likewise, we do not claim that these best fits are the global best fits. The results should not be treated as precise measurements of physical parameters. While they obviously hint at the TNBs true physical properties, these can only be used as starting points for future analyses.

Using our fits we can identify several broad trends. First, and most strikingly, TNBs with large primaries are more likely to have smaller best fit \jt{} values than small primaries, in line with theoretical predictions and observations of small bodies across the solar system. Another striking trend is that larger objects are more likely to have improvements using a non-Keplerian model, with the three greatest improvements (Salacia, Orcus, and Gonggong) being among the largest TNBs in our sample (which excludes Pluto, Eris, Haumea, and Makemake). This is likely because TNBs with large primaries tend to have shorter orbital periods (and thus stronger and faster non-Keplerian effects) in addition to having more high quality data. 

Another interesting discovery is a group of objects with orbit fit improvements at relatively high \jt{} (Borasisi, Altjira, 1999 RT$_{214}$). The high \jt{} indicated by our exploratory fits potentially hint that one of the system components may be a contact binary. While this is an unusual configuration, among asteroid binaries several primaries have been found with extreme values for \jt, perhaps most notably Kleopatra, with $J_2 = 0.765$ \citep[][]{brovz2021advanced}. Their prevalence in our findings, however, may be attributable to selection bias. Since these systems may have large \jt{} values, their non-Keplerian effects would be far easier to detect, when compared to objects with similar data quality, quantity, and observational baseline. As such, these systems may not be as common as would be suggested by our findings. 

Similar to a possible discovery of contact binaries, our results show three TNBs with unresolved binary-like \jt{} values (2005 EO$_{304}$, Teharonhiawako (``QT297''), and 2006 BR$_{284}$), although none of the fits for these bodies cross our improvement threshold. As discussed above, hierarchical triple systems may be sensitive tracers of planetesimal formation in the early solar system, making a discovery of additional hierarchical triple systems an important goal. In fact, \citet{nesvorny2021binary} conclude that Lempo-like triple systems ``should be found in the Kuiper Belt when observations reach the threshold sensitivity.'' More observations of these bodies, and subsequent orbit reanalysis, are required to find if these systems can be confirmed as having detectable non-Keplerian effects. Preliminary investigations suggest that these hierarchical triples are not resolvable with imaging. 

Another possible explanation for the large \jt{} values found for these three objects is that these TNBs are strongly affected by the Sun's gravitational influence, with this influence manifesting as a large measured \jt. All three are ultra-wide binaries, a class of TNBs with extremely large separation. Ultra-wides are most affected by the Solar tide due to their extremely long orbital periods. Future modeling efforts should aim to include the effects of the Solar tide to more fully model all gravitational dynamics at play. 

\begin{figure}
    \includegraphics[width=\linewidth]{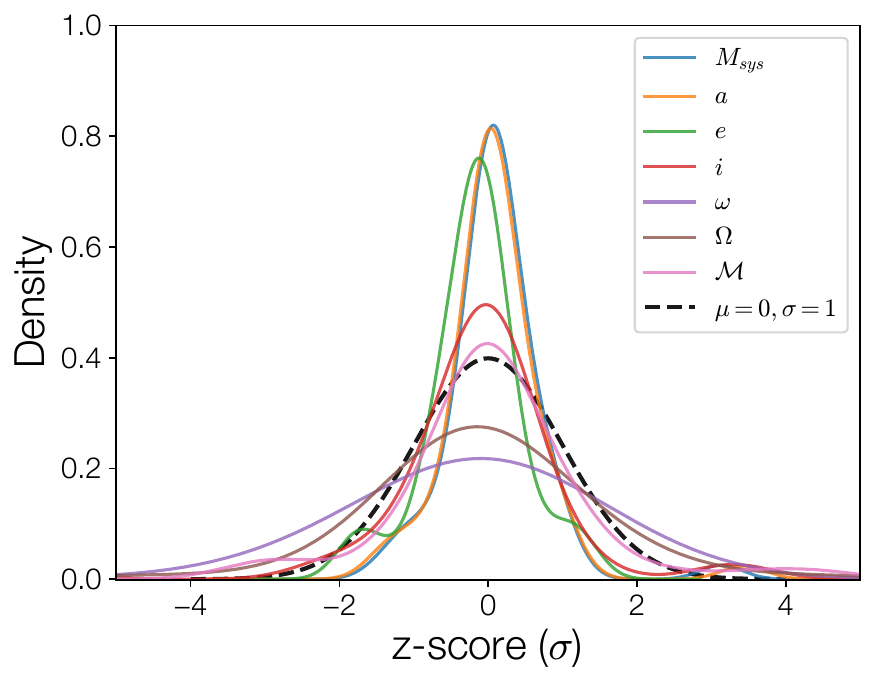}
    \caption{Kernel density estimates in the style of Figure \ref{fig:grundycomp}. In this figure, we compare the non-Keplerian best fits found in our exploratory fits to the Keplerian posteriors presented in Section \ref{sec:keplerianfits}. We find that significant deviations are present in the argument of periapsis ($\omega$) and longitude of ascending node ($\Omega$) when assuming a non-Keplerian model. This is expected since this model allows for precession of these two angles. The distributions are still centered at zero presumably because we are comparing the non-Keplerian angles at an epoch typically near the middle of the observations and because many systems show insignificant non-Keplerian effects. Notably, however, the system mass, semi-major axis, eccentricity, and inclination are still consistent with Keplerian fits, implying that systematic errors due to non-Keplerian effects are small in most previous analyses.}
    \label{fig:j2-comp}
\end{figure}

Our ensemble of exploratory fits is also able to identify systematic errors which may stem from the use of Keplerian fits in a variety of past analyses. To do this, it is most useful to compare our non-Keplerian best fits with the full Keplerian posteriors discussed in Section \ref{sec:keplerianfits}. Using the same methods as the comparisons to the Grundy best fits, we compare our fits in Figure \ref{fig:j2-comp}, using our Keplerian posteriors as the reference distribution. As can be seen, our non-Keplerian best fits are somewhat different to our Keplerian posteriors, especially for $\omega$ and $\Omega$. This is expected for non-Keplerian analyses since the addition of \jt{} shape effects allows for precession of these orbital angles. The absolute difference in these angles (at epoch), which can be several standard deviations from the Keplerian fit, are usually only a few degrees. 

Notably, the masses, semi-major axes, and eccentricities found in our non-Keplerian best fits are consistent with the Keplerian posteriors. This shows that systematic errors in orbital orientations can occur when assuming Keplerian orbits, but that masses, semi-major axes, periods, eccentricities, and inclinations are not significantly affected. Thankfully, the large number of analyses that rely on Keplerian fits are unaffected by systematic errors induced by the exclusion of non-Keplerian effects.  

\subsection{Identified Targets}

Here, we discuss in turn each of the eight TNBs identified as having detectable non-Keplerian effects. 

\subsubsection{(120347) Salacia-Actaea}
Salacia-Actaea is one of the largest and most massive TNBs in our sample. Our fits used 14 individual observations over a 10 year time span. Detection of non-Keplerian effects is expected as previous analyses showed that observations were 3.7$\sigma$ inconsistent with a Keplerian orbit \citep[][]{grundy2019mutual}. 

Our analysis had a best fit $J_2 = 0.0195$, which corresponds to a slightly oblate spheroid, but fits with similar likelihoods (although slightly worse) were different by a factor of a few. This large range of possible values for \jt{} is expected for systems where only apsidal or nodal precession is detected. In this case, Salacia's \jt{} is degenerate with its obliquity with respect to the mutual orbit. Since Salacia-Actaea's mutual orbit is nearly circular ($e = 0.0062^{+0.0031}_{-0.0027}$, Table \ref{tab:keplerian}) the non-Keplerian effects present in the system are likely to be nodal precession, since apsidal precession is difficult to detect at low $e$. Detection of nodal precession implies that Salacia's rotation pole is misaligned with the binary's mutual orbit, but the degree of misalignment is not possible to determine without further analysis. 

Future observations of the Salacia-Actaea system may be able to detect apsidal precession, allowing for a more full determination of the system's properties. Additionally, constraints provided by Salacia's low amplitude light curve \citep[][]{thirouin2014rotational} and future occultations may aid in this effort. 

\subsubsection{(90482) Orcus-Vanth}
Orcus-Vanth is another large, high mass TNB. Our fits used 12 individual observations taken over an almost 10 years. Like Salacia-Actaea, past analyses showed Orcus-Vanth was 2.2$\sigma$ inconsistent with a Keplerian orbit \citep[][]{grundy2019mutual}, making detection of non-Keplerian effects unsurprising. 

Our analysis had a best fit $J_2 = 0.016$, corresponding to a slightly oblate spheroid. Similar to Salacia-Actaea, due to Orcus-Vanth's nearly circular orbit only nodal precession was detected, introducing large uncertainties on our \jt{} and obliquity measurements. 

In the future, observations of Orcus-Vanth may help understand the system better, however, our preliminary results show that detection of apsidal precession may be an order of magnitude more difficult than for Salacia-Actaea. Understanding the shape of Orcus (or Vanth) will probably require additional multi-chord observations of stellar occultations, similar to the 2017 occultation of Vanth \citep[][]{sickafoose2019stellar}. 

\subsubsection{(225088) Gonggong-Xiangliu}
Gonggong-Xiangliu, the most massive binary in our sample, is the most surprising target with detection of non-Keplerian shape effects. Our fits only used 6 individual observations over about 8 years. The dearth of data, especially when compared to other TNBs, is due to the discovery of the moon in 2016 \citep[][]{kiss2017discovery}, although subsequent reanalysis found the moon in 2009 and 2010 HST images. Our detection is only robust when considering the prograde orbit solution. 

Our analysis had a best fit of $J_2 = 0.084$, corresponding to a fairly oblate spheroid. The high \jt{} we find is quite unexpected, and may possibly contradict light curve measurements \citep[][]{pal2016large}. In our analysis, we believe we have detected both apsidal and nodal precession, allowing us to place constraints on Gonggong's obliquity. Our best fit obliquity is 116.4$\degr$, however, our fits cannot distinguish between north and south pole. By folding the obliquities of our best fit solutions (the best fit and those with similar likelihoods) we find obliquities between 55$\degr$ and 65$\degr$ are favored. The large obliquity is surprising, but the binary's mutual orbit is also quite eccentric ($e = 0.2852^{+0.0086}_{-0.0079}$), possibly implying a complicated history of unusual tidal pumping \citep[][]{kiss2019mass,arakawa2021tidal}. We note that Gonggong's density has significant uncertainty that can be resolved when the obliquity of the satellite is clearly detected \citep{kiss2019mass}, making Gonggong-Xiangliu an extremely promising observational target.

Based on the small dataset, another possible explanation for this detection is overfitting by our model. With small amounts of data, model fitting can become more susceptible to overfitting. In general, 5 or more observations are required to fit a Keplerian orbit \citep[][]{grundy200842355}, so it seems suspicious that we find such high confidence detections of non-Keplerian effects, especially with no previously detected deviation from a Keplerian orbits (like Salacia and Orcus). As such, we do not place much confidence in this detection and consider the detection potentially spurious. 

To better understand Gonggong-Xiangliu's unexpected eccentricity and obliquity, resolve the mirror ambiguity, and/or understand issues with overfitting, additional observations need to be taken of this interesting system. The large difference in brightness between Gonggong and Xiangliu make these observations difficult without using the most advanced telescopes (e.g., HST, JWST, or possibly Keck). 

\subsubsection{(66652) Borasisi-Pabu}
Borasisi-Pabu, a small near-equal size binary in the Cold Classical Kuiper belt, was fit using 9 observations taken over about 8 years. No previous indications of poorly fitting Keplerian orbits have been reported in the literature, but our Keplerian analysis, which was consistent with past analyses, showed a Keplerian fit with a reduced $\chi^2$ of 1.9, indicating a somewhat poor fit to the data. 

We find a best fit $J_2 = 0.279$, consistent with an extremely elongated object, possibly similar to a contact binary. We also found an obliquity of $\sim$50$\degr$ for Borasisi, after detecting both apsidal and nodal precession. For Borasisi, the detectability of apsidal precession was enhanced by the mutual orbit's large eccentricity ($e = 0.4696^{+0.0017}_{-0.0017}$). 

We discuss Borasisi-Pabu in detail in Section \ref{sec:borasisi} and present fully converged, high quality non-Keplerian fits.

\subsubsection{(148780) Altjira}
Altjira (and its unnamed secondary), a small equal-size binary just outside the Cold Classical belt, was fit with 8 observations taken over about 4 years. No past analysis indicated poor quality Keplerian fits, but our Keplerian analysis indicated a marginally bad fit (reduced $\chi^2$ of 1.4). 

Our analysis resulted in a best fit $J_2 = 0.479$, roughly consistent with a contact binary. The measured \jt{} may also be caused by an unresolved system component, possibly making this a hierarchical triple system. We are able to constrain Altjira's obliquity to be between $\sim30\degr$ and $60\degr$ by detecting both apsidal and nodal precession. 

With $V = 23$ mag, little more is known about this system, with no known light curve or occultation measurements. Any additional constraints on this system will contribute to understanding the configuration of this system.  

\subsubsection{(174567) Varda-Ilmarë}
Varda-Ilmarë is a somewhat high mass TNB; however, unlike the other high mass TNBs with detected non-Keplerian motion, the primary-to-secondary brightness ratio is far more equal, putting it in a class that is somewhat intermediate between the low-mass (with near-equal masses) and high-mass (large primary-to-secondary mass ratio) TNBs. Our analysis used 12 observations taken over about 4 years. Previous Keplerian analyses did not indicate a significant deviation from Keplerian motion, but our Keplerian analysis showed that Keplerian orbits were somewhat poor quality fits for both the prograde and retrograde orbital solutions at the $\sim$1$\sigma$ level. Like with Gonggong-Xiangliu, our detection of non-Keplerian effects is significant for only the prograde orbital solution, although the retrograde solution almost reaches our improvement threshold. 

Our fits find a best fit $J_2 = 0.07$, consistent with a significantly flattened body. As with other large TNBs with near-circular orbits, we have only robustly detected nodal precession, leaving a degeneracy between \jt{} and obliquity..

Varda-Ilmarë is somewhat unique in having a mirror ambiguous orbit given the large quantity of high-quality data. This is primarily caused by the system's orbit which is viewed nearly face-on. Understanding Varda's role as a possible transitional system, may enable better understanding of the internal structure and composition of TNOs. 

\subsubsection{1999 RT$_{214}$}
1999 RT$_{214}$, a small near-equal size TNB in the Cold Classical belt, was fit using 6 observations taken over about 10 years. Past analysis did not show a significant deviation from a Keplerian orbit, but our Keplerian analysis indicated an extremely poor fit (reduced $\chi^2$ of 3.2), although with so few data points, such a result is possible $\sim$1\% of the time. With this amount of data, a detection of non-Keplerian motion is surprising, but the long observational baseline allows a fairly robust detection.

Our best fit analysis for this system gave $J_2 = 0.436$, corresponding to a contact binary-like shape. Another possible explanation is that a system component is an unresolved binary, possibly implying that 1999 RT$_{214}$ is a hierarchical triple. A wide range of permissible obliquities are allowed by our exploratory fits, with a range of $\sim30-55\degr$ being most likely. Our fits imply that both nodal and apsidal precession are visible. 

Additional observations are needed to confirm our detection and to further characterize this system. These will have to be completed by HST, JWST, or large ground-based observatories since $V = 24$ mag for the system. Combined with new observations, high quality non-Keplerian fits may be able to enable a better understanding of this system. 

\subsubsection{(79360) Sila-Nunam}
Sila-Nunam, the largest near-equal mass binary in the Cold Classical belt, was fit using 20 individual observations taken over 11 years. Uniquely, six of those observations are mutual events, where one system component passes in front of the other, from our view on Earth, producing a measurable drop in light. No previous modeling, including our Keplerian fits, hinted that non-Keplerian effects were detectable in this system, although previous measurements of Sila-Nunam's combined light curve showed that one, or both, components were flattened by $\sim$12\% \citep[][]{rabinowitz2014rotational}. 

Our fits find a best fit $J_2 = 0.199$, consistent with a considerably flattened body. In this case, since both components are possibly flattened, it is possible that both bodies contribute significantly to the overall system's \jt. Roughly speaking, a system where both bodies have a $J_2 \sim 0.1$ would match our results. Our exploratory fits cannot confidently determine whether nodal or apsidal precession (or both) are detected with high confidence, especially since the geometry of the system introduces several degeneracies in the model parameters. For example, $\omega$ and $\mathcal{M}$ had solutions that were good fits 180$\degr$ away from our best fit. These issues resulted in quality issues in our fits. Future non-Keplerian fits may be helped by introducing priors or reparameterizations which reduce the effects of degeneracy. 

In a tidally evolved system where the primary and secondary are locked in a spin-orbit resonance, like Sila-Nunam is thought to be, \ct{} can play a prominant role in the system's dynamics \citep[][]{proudfoot2021prolate}. This tidally evolved state is at odds, however, with the small, but potentially non-zero, eccentricity measured in a variety of analyses \citep[][]{grundy2012mutual,benecchi2014ut}, including both our Keplerian and non-Keplerian fits ($e = 0.0158^{+0.0167}_{-0.0098}$). This may be the result of relatively recent excitation which has yet to be damped out, or may be caused by unusual tidal dynamics.

Another explanation for the improvement in fit is overfitting. Sila-Nunam is unique in its use of mutual event data. Using mutual event data to aid in orbit fitting is unusual, but has been used in previous studies of this system without issue \citep[][]{benecchi2014ut}. We do, however, acknowledge that this data may introduce unforeseen systematic errors into our analysis, which may somehow result in better fits with a non-Keplerian model. Additionally, we found that a Keplerian model was a reasonably adequate fit to the data (reduced $\chi^2 \approx 1$), calling into question the need for a more complicated model.

The Sila-Nunam system is a prime target for future non-Keplerian analysis, especially with more complicated models that include the effects of \ct. This future analysis should also rigorously test if any systematic effects are introduced by the use of mutual event data and test whether overfitting may effect any resulting non-Keplerian fits. Given the mutual events and large set of photometric data, Sila-Nunam an ideal target for a thorough orbital analysis.

\subsection{Non-Keplerian Fitting of Other TNBs}
While the goal of this work is to identify the targets with the most detectable non-Keplerian shape effects, (and considerable work needs to be done to complete converged, high quality fits for these systems) other TNBs should not be ignored. Although the current data is not able to robustly detect non-Keplerian shape effects in systems not meeting our improvement threshold, full non-Keplerian fits can still be used to place upper limits on the presence of those effects. These upper limits may, in some cases, provide valuable constraints on the shapes of TNB components. 

To illustrate this, we consider our non-Keplerian fits to the 2002 UX$_{25}$ system. While our analysis only shows a slight improvement with the addition of a \jt{} and rotation pole, our fits show a rough upper limit on \jt{} of $\sim0.05$. While this rough estimate of an upper limit cannot be fully verified without converged non-Keplerian fits, it requires that 2002 UX$_{25}$ must have a remarkably spherical shape. When combined with its unusually small density \citep[$0.82 \pm 0.11$ g cm$^{-3}$,][]{brown2013density}, 2002 UX25 proves to be an unusual object and is possibly a touchstone for TNO formation \citep[e.g.][]{2019Icar..326...10B}. The low density suggests a significant porosity (potentially both macroporosity and microporosity) which seems unusual given its large size ($>$300 km radius), though central pressures are still relatively small as discussed by \citet{2019Icar..334...30G}. Additional formation modeling is needed to explain the presence of a small (collisionally-formed?) eccentric satellite around a large nearly-spherical rubble-pile. 


\subsection{Future Observations}
Based on our results, we believe that future observations should focus on those TNBs with detectable non-Keplerian effects (those described in the last section) and, possibly more importantly, TNBs that are near our improvement threshold. These TNBs show the most promise for future detections of non-Keplerian effects and provide the most efficient route for constraining the shapes of the largest number of objects. As observational baselines are extended, non-Keplerian effects become more detectable and uncertainties on the fits shrink. Observations can also be targeted at certain times when the uncertainty of the predicted position is largest, providing the most constraining observations (see Section \ref{sec:borasisi} for more details). 

In addition to high resolution imaging to resolve binary components, other types of observations, like light curves or occultations, still provide important orthogonal channels of information on the shapes and spins of TNBs. The constraints given by these observations are able to be meaningfully used as additional data sources in future more complicated models of TNB orbits. For example, the axial precession of a TNB primary due to back torques from the secondary, may be detectable in certain scenarios, which may enable a direct measurement of the primary-secondary mass ratio. Likewise, precession of the axes of TNO moons may provide information about the shapes and rotation poles of those moons \citep[as suggested by][]{hastings2016short}. 

\section{Borasisi-Pabu}
\label{sec:borasisi}

\begin{figure*}
\centering
\includegraphics[width=\textwidth]{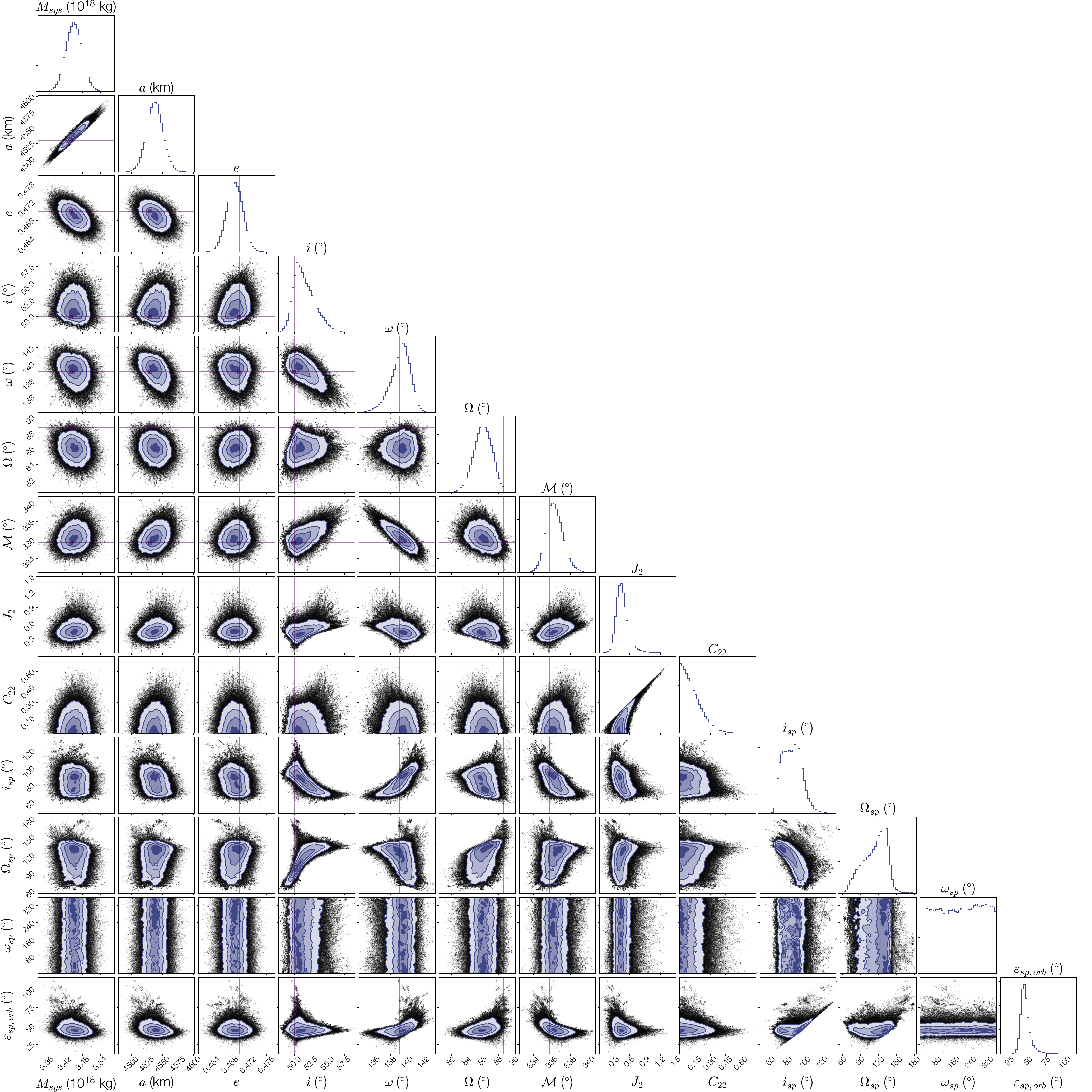}
\caption{Corner plot for the Borasisi-Pabu non-Keplerian orbit fit in the style of Figure \ref{fig:borasisi-kep}. Horizontal and vertical lines correspond to the best fit solution in the Grundy database. The best fit in the \multimoon{} non-Keplerian fit corresponded to a $\chi^2$ of $\sim$11.5 (compared to 21 for the Keplerian fit). 
Instead of individually plotting the primary and secondary masses, which are completely degenerate, the system mass ($M_{sys}$) is plotted to better show parameter correlations. Additionally, to allow for comparison with the literature, we show \jt{} and \ct{} rather than $J_2R^2$ and $C_{22}R^2$, where we take the shape to be a triaxial ellipsoid with $R$ as the volumetric radius. Of particular interest is the marginal distribution for \jt{} that clearly shows a high value of \jt{} is preferred, with $J_2 = 0$ (equivalent to Keplerian motion) being excluded with a significance $>3\sigma$. }
\label{fig:borasisi-nonkep}
\end{figure*}

\begin{figure}
\centering
\includegraphics[width=\linewidth]{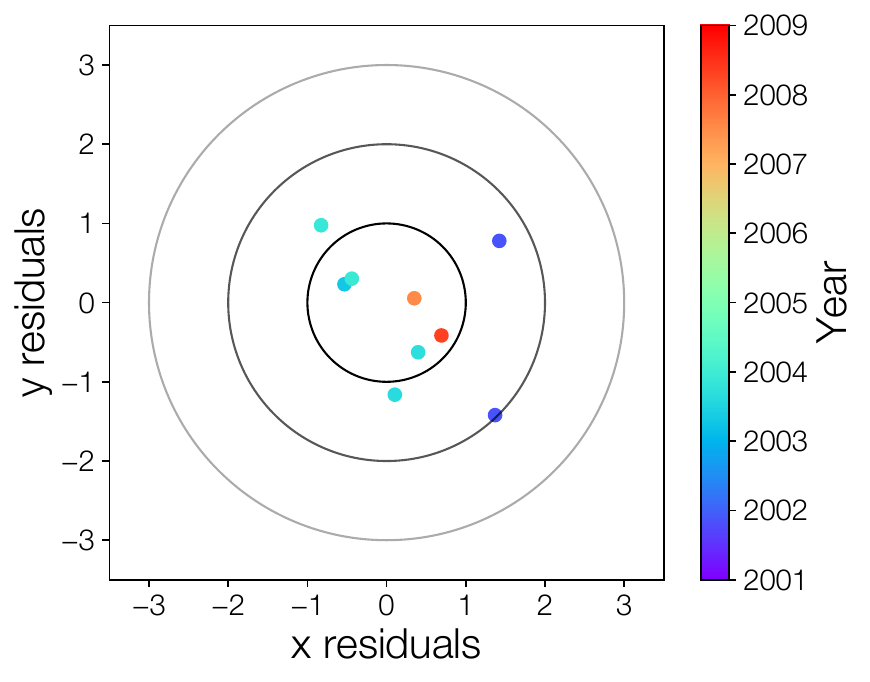}
\caption{Residual plot for the Borasisi-Pabu non-Keplerian orbit fit in the style of Figure \ref{fig:kep_residual}. The best fit in the \multimoon{} non-Keplerian fit corresponded to a $\chi^2$ of $\sim$11.5. RMS residuals are 2 milliarcseconds in both longitude and latitude. Error contours here are the same size as in Figure \ref{fig:kep_residual}.}
\label{fig:nonkep_residual}
\end{figure}

As a proof-of-concept we have performed detailed non-Keplerian fits to the orbit of Borasisi-Pabu, accounting for non-Keplerian effects due to both \jt{} and \ct. The marginal posterior distributions for each parameter are displayed in Table \ref{tab:borasisi} and joint posterior distributions are shown as a corner plot in Figure \ref{fig:borasisi-nonkep}. We also show a residual plot in Figure \ref{fig:nonkep_residual}.

\begin{deluxetable*}{lCCC}
\tablecaption{Non-Keplerian Orbit Solution for (66652) Borasisi-Pabu}
\tablehead{
Parameter &  & \text{Posterior Distribution} & \text{Best Fit}
}
\startdata
Fitted parameters & & & \\
\qquad Primary mass ($10^{18}$ kg) & M_1 & 2.293^{+0.667}_{-0.416} & 2.704 \\
\qquad Secondary mass ($10^{18}$ kg) & M_2 & 1.159^{+0.416}_{-0.668} & 0.747 \\
\qquad Semi-major axis (km) & a & 4538^{+13}_{-13} & 4535 \\
\qquad Eccentricity & e & 0.4690^{+0.0018}_{-0.0018} & 0.4694 \\
\qquad Inclination ($\degr$) & i & 51.285^{+1.920}_{-1.278} & 50.356 \\
\qquad Argument of periapse ($\degr$) & \omega & 139.31^{+1.03}_{-1.32} & 139.57\\
\qquad Longitude of the ascending node ($\degr$) & \Omega & 86.02^{+1.17}_{-1.20} & 85.90\\
\qquad Mean anomaly at epoch ($\degr$) & \mathcal{M} & 336.24^{+0.98}_{-0.85} & 336.00 \\
\qquad Primary zonal gravitational harmonic & J_2 & 0.4446^{+0.1192}_{-0.0928} & 0.3560 \\
\qquad Primary sectoral gravitational harmonic & C_{22} & 0.1033^{+0.1149}_{-0.0735} & 0.096 \\
\qquad Primary rotation axis obliquity ($\degr$) & i_{sp} & 84.85^{+11.30}_{-11.71} & 91.87 \\
\qquad Primary rotation axis precession ($\degr$) & \Omega_{sp} & 117.60^{+15.84}_{-25.85} & 98.75 \\
\qquad Primary rotation axis longitude ($\degr$) & \omega_{sp} & 184.40^{+120.10}_{-126.35} & 73.24 \\
Derived parameters & & & \\
\qquad System mass ($10^{18}$ kg) & M_{sys} & 3.451^{+0.029}_{-0.029} & 3.451 \\
\qquad Primary obliquity w.r.t. mutual orbit ($\degr$) & \varepsilon_{sp,orb} & 45.04^{+6.65}_{-4.66} & 43.15 \\
\qquad Orbit pole longitude ($\degr$) & \lambda_{orb} & 356.02^{+1.17}_{-1.20} & 355.90 \\
\qquad Orbit pole latitude ($\degr$) & \beta_{orb} & 38.72^{+1.28}_{-1.92} & 39.64 \\
\qquad Primary rotation pole longitude ($\degr$) & \lambda_{rot} & 27.60^{+15.84}_{-25.85} & 8.75 \\
\qquad Primary rotation pole latitude ($\degr$) & \beta_{rot} & 5.15^{+11.71}_{-11.30} & -1.87 \\
\hline
\enddata
\tablecomments{All fitted angles are relative to the J2000 ecliptic plane on JD 2451900.0 (2000 December 21 12:00 UT). Assumed c-axes for primary and secondary are 63 km and 54 km, respectively \citep[][]{vilenius2014tnos}. We use a rotation period of 6.4 hours \citep[][]{kern2006study}. Our fit presented here is only to the prograde obliquity solution, another solution with obliquity of $\sim135\degr$ also exists. Our fitted parameters $J_2R^2$ and $C_{22}R^2$ are presented here as \jt{} and \ct{} to make comparison with the literature easier; in this transformation we take $R$ to be volumetric radius. As mentioned in the text, we are not able to break the primary-secondary mass degeneracy, so we urge caution in using our fits for the masses. The best fit in the \multimoon{} non-Keplerian fit corresponded to a $\chi^2$ of $\sim$11.5 (compared to 21 for the Keplerian fit). We find that using our best fit Keplerian case, for this fit we find $\frac{\mathcal{L}_K}{\mathcal{L}_{NK}}$ = $\num{8.34e-3}$. RMS residuals are 2 milliarcseconds in both longitude and latitude.}
\label{tab:borasisi}
\end{deluxetable*}

Using the framework introduced in Section \ref{sec:nonkeplerian_fits}, we can assess the statistical significance of our non-Keplerian fit, when compared to the Keplerian fit from Section \ref{sec:keplerianfits} (see also Table \ref{tab:keplerian} and Figure \ref{fig:borasisi-kep}). As before, the Keplerian model we use is a subset of our non-Keplerian model (where $J_2$ and $C_{22}$ are 0). Using a likelihood-ratio test, we find that $\frac{\mathcal{L}_K}{\mathcal{L}_{NK}}$ = $\num{8.34e-3}$, giving confidence that the non-Keplerian model is statistically significant. This likelihood ratio statistic is somewhat smaller than that found in our exploratory fits (compare with Table \ref{tab:nonkeplerian}). Both Table \ref{tab:borasisi} and Figure \ref{fig:borasisi-nonkep} show that a Keplerian orbit ($J_2, C_{22} = 0$) for the mutual orbit of Borasisi-Pabu is strongly excluded, confirming that a non-Keplerian model is preferred. A non-Keplerian orbit model is also indicated by the poor quality Keplerian fit. Using the Keplerian fit described in Section \ref{sec:keplerianfits}, we find that there is a $\sim$3\% chance that random chance would result in a fit that is as bad (or worse) than the best fit we found. While not conclusive, this improvement merits further investigation.

To test how robust these results are to our input values (rotation period of Borasisi, c-axes of Borasisi and Pabu, integration tolerance), we also performed many additional non-Keplerian fits with these values substantially changed. These tests produced posteriors that are nearly identical to the one presented here, except when integration tolerance was set to a too high level. These tests showed that the choices we made for input values did not affect the fit presented here, and validate the inputs used in our exploratory fits. 

In addition to validating input parameters, we also ran a series of recovery tests on \multimoon. These tests used \multimoon{} to fit synthetically generated data (with artificially added uncertainties) using the best fit solution in Table \ref{tab:borasisi}, allowing us to determine if \multimoon{} could recover the input parameters. These tests returned similar fits to the one presented in this section, showing similar uncertainties and degeneracies, justifying the use of \multimoon{} given our current data and mitigating the risk of overfitting for this system.

For most parameters, there is agreement with both our Keplerian solution and the Grundy best fit, at the $\sim$1$\sigma$ level, except for $\Omega$, which shows a strong disagreement with Keplerian fits. This is an expected result of non-Keplerian fitting as the orbital model now allows for precession of the node. Our posterior shows $J_2 = 0.4446^{+0.1192}_{-0.0928}$ implying an extremely non-spherical shape of Borasisi. The spin pole direction is nearly perpendicular to the ecliptic. 

Our fits show no preference for a certain value of \ct, as can be seen in Table \ref{tab:borasisi} and the posterior distribution in Figure \ref{fig:borasisi-nonkep}. The posterior for \ct{} is purely a reflection of the prior limiting $C_{22} \leq \frac{1}{2}J_2R^2$. Since \ct{} is not detected, the corresponding spin longitude parameter, $\omega_{sp}$, is unconstrained, with a uniform distribution between 0-360$\degr$. This is expected since the mutual orbit is much longer than Borasisi's rotation period ($\sim$46 days and $\sim$20 hours, respectively). In future analyses, exclusion of \ct{} may be appropriate, when the orbit to spin period ratio is large. This may reduce the computational expense of the fitting process. 

In addition to constraining Borasisi's \jt{}, and by proxy, its shape, we can also place constraints on Borasisi's obliquity w.r.t. the mutual binary orbit. Our analysis shows an obliquity of $\varepsilon_{sp,orb} = 45.04^{+6.65}_{-4.66}$. This fit only reflects the prograde obliquity solution, where Borasisi's rotation pole and the binary's mutual orbit lie in the same hemisphere. Another obliquity solution (with $\varepsilon_{retrograde} = 180\degr - \varepsilon_{prograde}$) exists and provides a similar fit to the data. Distinguishing between these solutions may be possible with additional relative astrometry, depending on how the system's angular momentum is partitioned. In a future work, we will explore why two solutions exist and when they are distinguishable. 

Although we do present the individual mass posteriors in Table \ref{tab:borasisi}, these are highly degenerate in the current fits and should not be used individually. Prospects for breaking the mass degeneracy will be examined in a future work. 


The match between this full non-Keplerian fit and our exploratory fits supports our conclusion that the exploratory fits have identified promising targets for future non-Keplerian analysis and additional observations.

\subsection{Discussion on the inferred \jt{} of Borasisi}
Based on the large value of Borasisi's \jt{} from our analysis, several interpretations of the results are possible. First, and most simply, Borasisi is extremely flattened, and probably elongated, with a shape most likely resembling a contact binary. Contact binaries are common among the TNOs \citep[][]{thirouin2018plutino,thirouin2019light,showalter2021statistical} and can be part of TNB systems \citep[][]{rabinowitz2019complex}. With current constraints on the \jt, it is difficult to say more about Borasisi's shape, but with narrowed uncertainties, detailed shape modeling may be able to distinguish between snowman (like Arrokoth) or peanut contact binaries (like Kleopatra) \citep[as in][]{marchis2005mass}.

Our measurement of Borasisi's obliquity ($\sim45\degr$) is somewhat surprising, as most asteroid binaries have orbits fairly well-aligned with the rotation pole of the primary. TNBs, however, are formed differently and exhibit properties not seen in asteroid binaries. The closest comparison to Borasisi-Pabu is Manwë-Thorondor; Manwë is possibly a contact binary inclined to its mutual orbit by 27$\degr$ \citep[][]{rabinowitz2019complex}. Both are Cold Classicals. Combined with a similar mutual orbit eccentricity, total system mass, and separation, the Manwë-Thorondor system is strikingly similar to the Borasisi-Pabu system. Indeed, if we calculate Manwë's $J_2$ (based on the shape model from \citet{rabinowitz2019complex}), we find a value very similar to Borasisi's. These objects may be part of a larger population of contact (or compact) binaries with oblique satellites.

The study of this population of objects should focus on explaining how these objects formed and evolved over the history of the solar system. One possible hypothesis is that these systems formed with relatively circular and coplanar orbits. Then, due to their somewhat loose separations, close encounters with other TNOs (or TNBs) may have excited their mutual orbits \citep[similar to][]{campbell2022close}. Tidal dissipation is unlikely to be effective at damping eccentricity and inclination in these systems due to the system's large separation and \jt{} \citep[][]{porter2012kctf}, so any past excitation would remain visible in their orbits today. Alternatively, these systems may have formed with eccentric and inclined orbits. More advanced simulations of TNB formation, especially of gravitational collapse of pebble clouds, need to be completed before distinguishing between these (or other) possibilities. 

One shortcoming of our model, that may affect our interpretation, is the assumption that Pabu's shape does not contribute to the system dynamics. This is likely an oversimplification given Pabu's large relative size. However, assuming that Borasisi and Pabu have the same obliquities (w.r.t. the mutual orbit) and shapes, we can calculate the required \jt{} for each body to match our median $J_2R^2$, assuming the volumetric radius ratio of Borasisi and Pabu is $\sim$0.8 \citep[][]{vilenius2014tnos}. We find that each body would have $J_2 \sim 0.3$, still a shape that is consistent with a contact binary \citep[see Figure 9 of][]{marchis2005mass}. In the case where the obliquities are not aligned, \jt{} for each object would have to be even higher. This still implies that at least one system component must be a contact or compact binary. In the future, if more data becomes available for the Borasisi-Pabu system, a more complicated model of the system's dynamics including quadrupole gravity for Pabu may be warranted.

\begin{figure*}[ht]
\includegraphics[width=\textwidth]{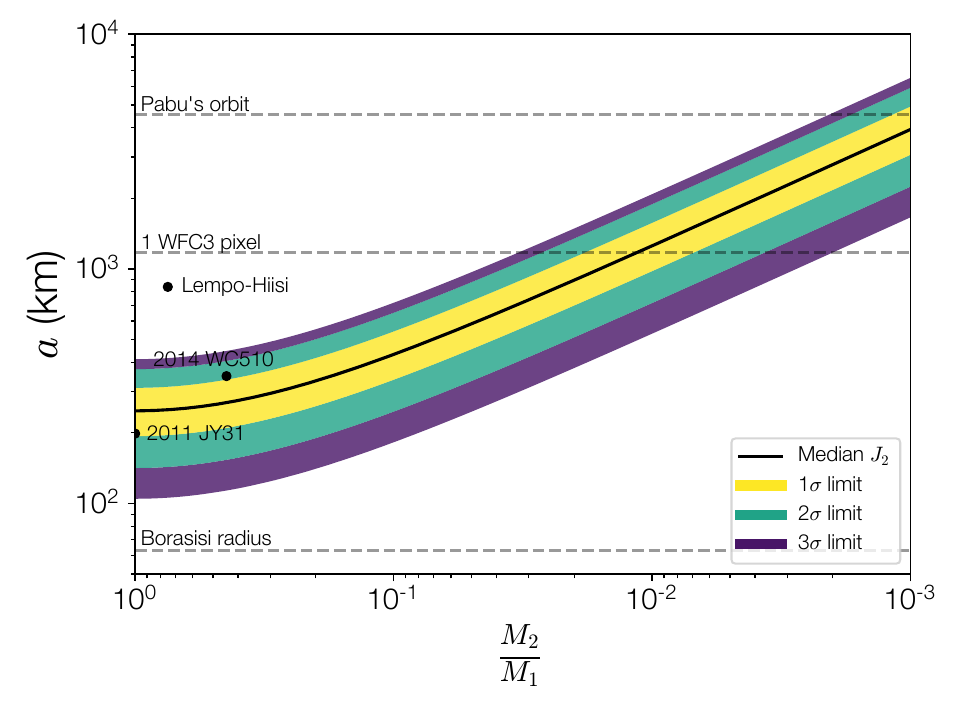}
\caption{Constraints on the mass-ratio and separation of a putative unresolved binary in the Borasisi system. The thick black line shows the mass ratio and separation given by the median value of the inferred \jt{} implied by equation \ref{eqn:j2binary}. Colored contours indicate the 1, 2, and 3$\sigma$ ranges from the \jt{} posterior. Gray dashed lines show several separations for reference. Here, we adopt a radius of 63 km for Borasisi, but this could be smaller, especially at mass ratios close to 1. The separation equivalent to a single Wide Field Camera 3 (WFC3) pixel on HST is calculated based on Borasisi-Pabu's typical distance from Earth and the instrument's 0.04'' px$^{-1}$ pixel scale. The separation of Pabu's orbit is shown in Table \ref{tab:borasisi}. The black points show other close binaries known in our solar system. Data for Lempo-Hiisi is from \citet{bpm1spinnymultimoon}, 2014 WC510 is from \citet{leiva2020stellar} and assumes equal densities and albedos, and for 2011 JY31 is from \citet{weaver2022high} and assumes equal masses. The \jt{} found in our analysis is consistent with an unresolved binary interpretation, as the separation of the binary would be less than a single HST pixel across a wide range of mass ratios.}
\label{fig:borasisi-binarysep}
\end{figure*}

A second possible interpretation of the large $J_2$ value found in our analysis is that the Borasisi-Pabu system hosts another unresolved component. The possible properties of this unresolved component, assuming all three components are spheres, can be explored using Equation \ref{eqn:j2binary}. In Figure \ref{fig:borasisi-binarysep}, we plot the possible configurations (in terms of mass ratio and separation) of the unresolved component based on our median $J_2R^2$ and its 1, 2, and 3$\sigma$ uncertainties. It clearly shows that a wide range of allowed system configurations exist. While some are potentially resolvable with HST's Wide Field Camera 3 (WFC3), it is unlikely that resolvable satellites (with separations of $\gtrsim$2000 km) would be dynamically stable for long times with Pabu (periapse of 2400 km). However, unresolvable near-equal-mass inner binaries with separations of a few hundred kilometers are quite possible. We also draw comparisons with other known tight binaries 2011 JY$_{31}$, 2014 WC$_{510}$, and Lempo-Hiisi, the inner binary in the Lempo triple system. These three comparisons have a range of masses, with Lempo being much more massive than Borasisi-Pabu and 2011 JY$_{31}$ being much less massive, but they illustrate that an unresolved system component is a plausible explanation for the detected non-Keplerian effects. 

Under the unresolved binary interpretation, the obliquity measured in our analysis is a proxy for the inclination of the unresolved object's orbit w.r.t. the orbit of Pabu. An inclination of $\sim45\degr$ is above the critical inclination where the Kozai-Lidov mechanism can be active ($i_{crit} \approx 39.2\degr$), but if the central components are significantly aspherical, Kozai-Lidov cycling can be avoided. The short- and long-term dynamics of triple systems are extremely interesting and complex, but we leave a full investigation of them to future work. 


Assessing the relative probability of each of our interpretations (contact binary and unresolved component) is impossible at the 2-quadrupole level. However, additional 3-body analysis with \multimoon{} may allow us to infer the orbit of an undetected system component since it would induce somewhat different non-Keplerian motion. Additionally, other observations of light curves, thermal emission, or occultations may enable us to distinguish between these two interpretations. 

After completing all the analysis discussed in this Section, a new study of TNO light curve found a different light curve period for Borasisi of $19.868 \pm 0.032$ hours with an amplitude of $\Delta m = 0.216 \pm 0.057$ \citep[][]{kecskemethy2023light}. However, our testing showed that our results were robust to different primary rotation periods, so we do not include this new rotation period in our analysis. The fairly high amplitude light curve, however, is generally compatible with our results.

\subsection{Possible Systematics}
Although our analysis is promising, there is still a distinct possibility that our results could be due to systematic errors. When comparing Figures \ref{fig:kep_residual} and \ref{fig:nonkep_residual}, it seems that the improvement in orbit fit is driven by just one data point near the end of the dataset. This is expected from a true non-Keplerian orbit, as non-Keplerian deviations grow linearly in time, but could also indicate unknown systematics. The next data point seems to be consistent with a Keplerian orbit, further raising concern for systematics. However, upon examination, that observation was taken near periapse, when the non-Keplerian signal would be the smallest. 

To check for possible systematics, we completed a fit without the second-to-last data point (from $\sim$2007). In this fit, we find that $\frac{\mathcal{L}_K}{\mathcal{L}_{NK}} \sim 0.3$, implying that the non-Keplerian fit shows only modest improvement over a Keplerian fit. Given the lack of data (just 9 observations), this is expected behavior for a true non-Keplerian orbit, but it still does not conclusively rule out possible systematics. At this time it is impossible to conclusively rule against (or for) systematic errors. We do point out however, that our non-Keplerian model is able to match the observations extremely well, which would be unlikely to occur given randomly distributed systematic errors. Additionally, our model outputs are physically realistic and similar to other known binary systems containing a contact binary (e.g., Manwë). Understanding if systematic errors can masquerade as non-Keplerian effects for the Borasisi-Pabu system will require additional observations of the system. We note that Borasisi, Sila, 1999 RT214, and Altjira are targets for HST Program 17206 (PI: Proudfoot) where two additional observations will be made to improve our ability to interpret non-Keplerian motion and systematic uncertainties.

In future investigations, sources of systematic errors should be thoroughly examined. These could come from a variety of sources, including: cosmic ray hits on images, inaccurate PSF models, uncertainties in the plate solution, under/overestimated astrometric uncertainties, center-of-light--center-of-mass offsets (see Appendix A), and  others. In the future, it may be useful to use a likelihood model that includes a flexible noise model so \multimoon{} can explicitly infer the noise properties of the astrometric measurements.

\begin{figure*}[t]
    \includegraphics[width=\textwidth]{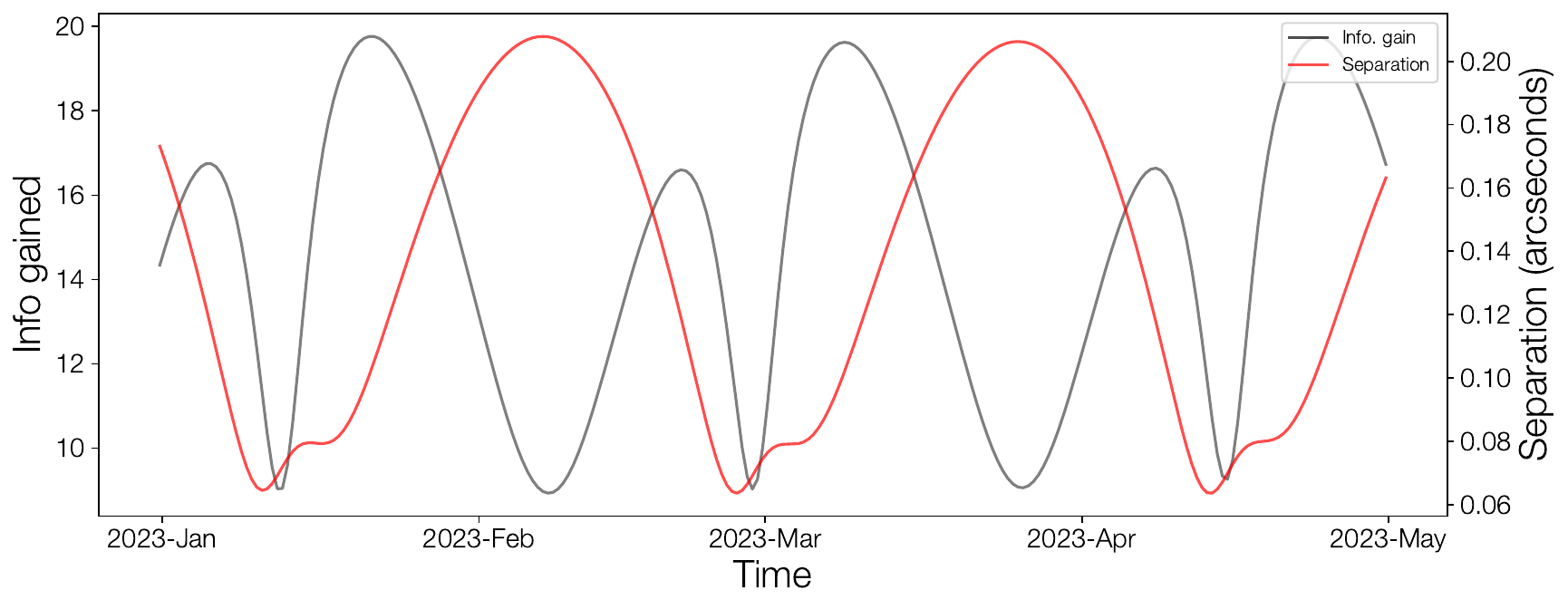}
    \caption{A plot showing the change in ``information gained'' by a single HST observation (in gray) and separation of Borasisi and Pabu (in red) in the first quarter of 2023. The information gained by single observations is a statistic which compares the uncertainty in the projected position of a system's secondary to the uncertainty of a typical observation (Equation \ref{eqn:ig}). Information gained clearly shows large variations; observing at times of maximum information gained allows for maximum leverage over reducing uncertainties. Additionally, avoiding times of close separation can ensure high quality data.}
    \label{fig:infogain}
\end{figure*}

\subsection{Potential For Future Observations}
Future observations are key to confirming the results we present here, and will enable us to significantly narrow the uncertainties associated with our modeling. Any additional observations will strengthen the detection of non-Keplerian effects, and may enable us to break the degeneracies associated with our modeling. To illustrate the ability of future observations to improve our orbital models, in Figure \ref{fig:infogain} we have plotted the ``information gain'' as a result of additional observations at any given time during the first quarter of 2023, along with the separation of the binary. We define information gain as the size of the cloud of projected positions of the secondary, in units of typical HST observations. This can be written:

\begin{equation}
\label{eqn:ig}
   I.G.(t) = \sqrt{\left( \frac{\sigma_{x,pred}(t)}{\sigma_{x,typ}} \right)^2 + \left( \frac{\sigma_{y,pred}(t)}{\sigma_{y,typ}} \right)^2}
\end{equation}

where $\sigma_{x,pred}$, $\sigma_{y,pred}$ are the standard deviations of the projected positions from a sample of posterior draws and $\sigma_{x,typ}$, $\sigma_{y,typ}$ are typical error measurements of Pabu's position, based on past HST observations of Borasisi-Pabu. To calculate this, we have taken 1000 random samples from our MCMC chain and used them to predict the position of Pabu at a given time. We then use the ensemble to calculate the information gain.

This is equivalent to calculating the average expected $\sqrt{\chi^2}$ of a new relative astrometric observation given the current posterior distribution (as listed in Table \ref{tab:borasisi}). For an observation with no additional constraining power, information gained is $\sqrt{2}$. Observations can be optimized to best shrink uncertainties by scheduling when information gain is at its maximum. In the case of Borasisi-Pabu, observations in 2023 can significantly shrink measurement uncertainties, which is unsurprising given that Borasisi-Pabu was last observed by HST in 2008. Targeting observations at times of high information gain allow for optimization of observing schedules. In addition, observing when Borasisi and Pabu have the greatest separation ensures high data quality, further motivating observations at certain times. 

\section{Conclusions}
\label{sec:conclusion}
Using a new non-Keplerian Bayesian orbit fitter, \multimoon, we have completed a set of Keplerian and non-Keplerian orbit fits to 45 TNBs. Our Keplerian orbit fits are in close agreement with those previously completed in the literature, validating \multimoon's orbit fitting procedures. Our exploratory non-Keplerian fits were run with the goal of identifying targets for full non-Keplerian analysis and possible observational campaigns. Almost 20\% (8/45) of TNBs analyzed have improvements in fit quality when using a non-Keplerian orbit model, with many systems with moderate improvement in fit quality. The 8 systems we identify for future analysis and observation are Salacia-Actaea, Orcus-Vanth, Gonggong-Xiangliu, Borasisi-Pabu, Altjira, Varda-Ilmarë, 1999 RT$_{214}$, and Sila-Nunam. 

Our analysis is consistent with previous expectations that large TNOs are nearly spherical. The large TNB systems with detectable non-Keplerian effects are all consistent with an oblate spheroidal shape. We find that small TNBs host extremely aspherical components. Of the small TNBs with detectable non-Keplerian effects, almost all are consistent with contact binary shapes, though we are biased to highly non-spherical objects. We also identify a possible population of widely separated hierarchical triples, although these detections individually do not quite reach our improvement threshold. Our ensemble of exploratory non-Keplerian fits is consistent with their Keplerian counterparts for the TNB system's mass, semi-major axis, eccentricity, and inclination. This indicates that there are no systematics introduced by non-Keplerian effects that may invalidate past Keplerian orbital analyses of TNBs. 

As an illustration of how our target list can be used to inform future non-Keplerian analyses, we have completed a full non-Keplerian fit to the mutual orbit of Borasisi-Pabu. Our analysis confirms the finding of our exploratory fits, finding that non-Keplerian motion is detectable in the system. We find that, assuming a triaxial shape, Borasisi (and/or Pabu) is extremely flattened and elongated, and may be a contact binary. Another interpretation of our results is that Borasisi (and/or Pabu) may themselves be a compact binary, making the system a hierarchical triple system like Lempo. Alternatively, our results may be caused by unidentified systematic errors present in a small fraction of our dataset. In any case, future observations are necessary to fully understand the system's architecture.

Future observations of potentially non-Keplerian TNBs are of high priority since these effects are a unique probe into TNB shapes and sizes. Many of the systems we have identified as having currently detectable non-Keplerian effects have not been observed in more than a decade allowing significant orbital precession to accumulate. High resolution relative astrometry combined with non-Keplerian models can further confirm our results, narrow uncertainties, and break degeneracies. 

\begin{acknowledgments}

We thank Simon Porter and Seth Pincock for help with \texttt{SPINNY}. We also thank Mayalen Laker as a Physics 227 Teaching Assistant. 

This research is based on observations made with the NASA/ESA Hubble Space Telescope obtained from the Space Telescope Science Institute, which is operated by the Association of Universities for Research in Astronomy, Inc., under NASA contract NAS 5–26555. The work completed here were funded by programs HST-AR-14581 and HST-GO-15460. This work was also supported by grant 80NSSC19K0028 from NASA Solar System Workings. 

\end{acknowledgments}

\appendix
\restartappendixnumbering

\section{Keplerian Fits with Constant Offsets}
\label{sec:kepoff_methods}
In addition to regular Keplerian fits, we completed another set of Keplerian fits in which the model allows for a constant offset between a TNB primary's center of light (photocenter) and center of mass. When observing TNOs, photocenter offsets can be caused by three different mechanisms. Firstly, at non-zero phase angles, the photocenter and center of mass of an object will be slightly misaligned, even for a spherical object, due to the viewing geometry. This effect is enhanced when objects are aspherical or have drastic albedo features (e.g., Sputnik Planitia on Pluto). The misalignment, however, even for extremely aspherical shapes and extreme albedo variations will generally only amount to a small fraction of the object's overall angular diameter. The angular diameters of the TNB components in our dataset are $\lesssim$40 milliarcseconds, therefore, constant misalignments between the photocenter and center of mass of an object will be $\lesssim$5 milliarcseconds for the largest objects in our sample, and much smaller for the majority of objects. This is comparable to the observational uncertainty of a few milliarcseconds for most objects. 

Secondly, previously undetected close-in satellites can also cause photocenter offsets, due to light contamination from the secondary. For single observations the offset could be sizeable, depending on the mass, density, and albedo of the undetected satellite, but given a series of observations, the misalignment from a close-in satellite on a circular orbit will generally average out to zero. For an eccentric satellite however, the satellite will spend more time near apoapse, naturally causing an average photocenter-barycenter misalignment. Incomplete sampling of the putative undetected satellite's orbit and an eccentric orbit can both cause an average change in the primary's photocenter.

Third, objects with complex shapes and potentially heterogeneous interiors can have center-of-mass--center-of-figure offsets. Even neglecting the above effects, this would lead to a center-of-light offset. 

A full model of potential center-of-light/center-of-mass/center-of-figure variations for non-spherical shapes under varying illumination and observation conditions would be highly complex. Most of this complexity would be well below the detectability threshold. However, to explore these effects we add two parameters to our Keplerian model that represent a simple constant offset to the modeled astrometry (in ecliptic longitude and latitude). These fits allow us to consider any potential astrometric offsets produced by aspherical objects (e.g., the center of reflected light from an object may be offset from the center of mass) or contamination from previously undetected components of the TNB system (e.g., small moons close to the primary that are unresolvable with current telescopes). These fits were run and analyzed identically to the normal Keplerian fits. 

In our ensemble of Keplerian fits with constant offsets, results were varied. In most cases, the fits, when compared to the normal Keplerian fits, showed little to no improvement in quality. Despite this, a few objects had large improvements to the quality of the fits. However, the majority of these large improvements came at the cost of unrealistically large constant offsets (no significant improvements were obtained with offsets less than 5 mas). While this may indicate the presence of undetected components in the TNB system, many of the offsets were probably much too large to be caused by plausible undetected system components (offsets could be as large as $\sim$90 mas).

The improvement in fit is most likely due to overfitting. Our parametrization will always enable an improvement in fit by shifting the 2-dimensional residual cloud towards the origin. This is amplified by the small dataset used, with some TNBs only having 5 observations. This, combined with the physically unrealistic offsets found, convince us that the use of constant photocenter offsets is not warranted at this time. 

Although we discard these fits, some insights can still be learned from examining their results (which are available upon request). In cases where there is improvement to the quality of the fit over the Keplerian (small or large improvements), the inclusion of constant offsets generally alters the fitted eccentricity of the TNB's orbit. In most cases, the fits with offsets show a small decrease in the fitted eccentricity, although increases can occur. When the Grundy best fits (without offsets) are compared to the posterior distributions from the constant offset fits (in a similar way to the data in Figure \ref{fig:grundycomp}), the \textit{z}-scores are similarly small for all parameters except eccentricity. This indicates that even if constant photocenter offsets are present in TNB relative astrometry, results based on the ensemble of TNB orbits (e.g., the mutual inclination distribution) are unaffected.

The non-detection of any believable constant photocenter offset, is not overly surprising. In reality, photocenter offsets (from any source) are time-varying. For offsets due to viewing geometry, most of the variation is caused by the Earth's orbit around the Sun, with the TNB's heliocentric orbit making a small contribution. Offsets from albedo variations and aspherical shapes will vary with the rotational period of the object in question. Future investigations of photocenter offsets should use a more physically realistic model, which accounts for the phase angle and rotation phase/rate of the modeled objects. These models may be able to constrain the shapes and albedo variations of TNB components once sufficiently high resolution data is available. 

\longrotatetable
\begin{deluxetable}{lcccccccc}
\tabletypesize{\footnotesize}
\tablewidth{\textwidth}
\tablecaption{Keplerian Orbital Fits}
\tablewidth{0pt}
\tablehead{
Object & RMS residual & $M_{sys}$ & $a$ (km) & $e$ & $i$ & $\omega$ & $\Omega$ & $\mathcal{M}$ \\
 & (arcsec) & ($10^{18}$ kg) & (km) & & ($^{\circ}$) & ($^{\circ}$) & ($^{\circ}$) & ($^{\circ}$)
}
\startdata
(26308) 1998 SM$_{165}$ & 0.001 & $6.880^{+0.015}_{-0.015}$ & $11372^{+8}_{-8}$ & $0.4733^{+0.0006}_{-0.0006}$ & $66.303^{+0.061}_{-0.062}$ & $49.32^{+0.11}_{-0.11}$ & $52.35^{+0.07}_{-0.07}$ & $165.51^{+0.09}_{-0.09}$ \\ 
(42355) Typhon-Echidna & 0.001 & $0.946^{+0.053}_{-0.05}$ & $1625^{+30}_{-29}$ & $0.5259^{+0.0155}_{-0.0149}$ & $49.563^{+1.935}_{-1.785}$ & $129.51^{+2.09}_{-2.06}$ & $230.55^{+3.41}_{-3.32}$ & $19.08^{+1.76}_{-1.81}$ \\
(50000) Quaoar-Weywot, \textit{pro.} & 0.004 & $1210.333^{+20.811}_{-20.708}$ & $13315^{+76}_{-76}$ & $0.0217^{+0.0055}_{-0.0052}$ & $14.636^{+0.501}_{-0.493}$ & $298.49^{+9.14}_{-8.50}$ & $353.16^{+1.43}_{-1.39}$ & $236.77^{+8.34}_{-8.97}$ \\
(50000) Quaoar-Weywot, \textit{retro.} & 0.004 & $1215.502^{+22.613}_{-21.104}$ & $13335^{+82}_{-78}$ & $0.0217^{+0.0052}_{-0.0057}$ & $150.011^{+0.521}_{-0.514}$ & $298.45^{+9.08}_{-8.97}$ & $346.44^{+0.78}_{-0.71}$ & $241.51^{+8.70}_{-8.77}$ \\
(55637) 2002 UX$_{25}$ & 0.009 & $124.615^{+3.010}_{-2.717}$ & $4771^{+38}_{-35}$ & $0.1735^{+0.0237}_{-0.0243}$ & $84.447^{+0.265}_{-0.278}$ & $16.52^{+1.22}_{-1.41}$ & $203.30^{+0.27}_{-0.27}$ & $289.39^{+1.43}_{-1.33}$ \\
(58534) Logos-Zoe & 0.005 & $0.452^{+0.009}_{-0.008}$ & $8180^{+55}_{-53}$ & $0.5424^{+0.0073}_{-0.0073}$ & $72.893^{+0.558}_{-0.563}$ & $201.47^{+0.53}_{-0.52}$ & $17.06^{+0.39}_{-0.40}$ & $35.35^{+0.87}_{-0.90}$ \\
(60458) 2000 CM$_{114}$ & 0.003 & $2.018^{+0.053}_{-0.054}$ & $2504^{+22}_{-23}$ & $0.0389^{+0.0150}_{-0.0140}$ & $71.032^{+1.894}_{-1.763}$ & $215^{+20}_{-17}$ & $278.34^{+2.26}_{-2.40}$ & $71^{+280}_{-63}$ \\
(65489) Ceto-Phorcus, \textit{pro.} & 0.001 & $5.423^{+0.338}_{-0.364}$ & $1842^{+38}_{-42}$ & $0.0095^{+0.0066}_{-0.0049}$ & $76.238^{+3.135}_{-2.464}$ & $251^{+57}_{-151}$ & $112.78^{+3.53}_{-3.32}$ & $244^{+89}_{-185}$ \\
(65489) Ceto-Phorcus, \textit{retro.} & 0.002 & $5.687^{+0.352}_{-0.339}$ & $1872^{+38}_{-38}$ & $0.0113^{+0.0080}_{-0.0057}$ & $133.708^{+2.312}_{-2.513}$ & $281^{+36}_{-141}$ & $121.05^{+4.36}_{-4.27}$ & $301^{+35}_{-197}$ \\ 
(66652) Borasisi-Pabu & 0.002 & $3.432^{+0.026}_{-0.027}$ & $4528^{+12}_{-12}$ & $0.4696^{+0.0017}_{-0.0017}$ & $50.004^{+0.318}_{-0.315}$ & $139.24^{+0.31}_{-0.31}$ & $88.68^{+0.48}_{-0.48}$ & $335.58^{+0.46}_{-0.46}$ \\
(79360) Sila-Nunam & 0.003 & $10.821^{+0.206}_{-0.199}$ & $2775^{+17}_{-17}$ & $0.0158^{+0.0167}_{-0.0098}$ & $120.829^{+0.339}_{-0.342}$ & $173^{+14}_{-112}$ & $134.00^{+0.63}_{-0.63}$ & $50^{+149}_{-14}$ \\
(80806) 2000 CM$_{105}$ & 0.006 & $0.823^{+0.079}_{-0.085}$ & $2348^{+72}_{-83}$ & $0.0228^{+0.0158}_{-0.0112}$ & $61.613^{+3.037}_{-3.087}$ & $244^{+30}_{-85}$ & $201.71^{+3.21}_{-2.65}$ & $257^{+47}_{-151}$ \\ 
(88611) Teharonhiawako-Sawiskera & 0.027 & $2.437^{+0.027}_{-0.027}$ & $27645^{+103}_{-103}$ & $0.2484^{+0.0018}_{-0.0018}$ & $127.637^{+0.332}_{-0.336}$ & $301.84^{+0.49}_{-0.49}$ & $36.43^{+0.46}_{-0.46}$ & $276.05^{+0.44}_{-0.45}$ \\
(90482) Orcus-Vanth & 0.003 & $632.995^{+3.468}_{-3.728}$ & $8991^{+16}_{-18}$ & $0.0012^{+0.0007}_{-0.0006}$ & $90.276^{+0.344}_{-0.329}$ & $250^{+37}_{-37}$ & $50.94^{+0.37}_{-0.40}$ & $222^{+33}_{-44}$ \\
(119979) 2002 WC$_{19}$, \textit{pro.} & 0.010 & $83.053^{+5.464}_{-5.465}$ & $4198^{+90}_{-94}$ & $0.2357^{+0.0372}_{-0.0401}$ & $25.482^{+0.667}_{-0.625}$ & $137.35^{+2.51}_{-2.76}$ & $120.59^{+1.92}_{-1.89}$ & $354.29^{+4.29}_{-352.84}$ \\
(119979) 2002 WC$_{19}$, \textit{retro.} & 0.010 & $70.926^{+3.714}_{-3.342}$ & $3983^{+68}_{-64}$ & $0.1955^{+0.0428}_{-0.0434}$ & $160.106^{+0.763}_{-0.765}$ & $154.51^{+3.29}_{-4.06}$ & $232.19^{+2.20}_{-2.19}$ & $350.46^{+3.64}_{-3.73}$ \\ 
(120347) Salacia-Actaea & 0.011 & $488.855^{+10.003}_{-9.911}$ & $5710^{+39}_{-39}$ & $0.0062^{+0.0031}_{-0.0027}$ & $17.905^{+0.515}_{-0.493}$ & $257^{+80}_{-239}$ & $109.97^{+1.75}_{-1.84}$ & $197^{+63}_{-38}$ \\
(123509) 2000 WK$_{183}$, \textit{blended} & 0.003 & $1.100^{+0.037}_{-0.032}$ & $2367^{+26}_{-23}$ & $0.0132^{+0.0070}_{-0.0059}$ & $93.8^{+7.4}_{-9.9}$ & $200^{+35}_{-56}$ & $158^{+17}_{-8}$ & $260^{+51}_{-38}$ \\
(134860) 2000 OJ$_{67}$, \textit{blended} & 0.002 & $1.908^{+0.151}_{-0.118}$ & $2271^{+58}_{-48}$ & $0.0354^{+0.0346}_{-0.0188}$ & $95.1^{+10.8}_{-11.3}$ & $147^{+89}_{-32}$ & $232^{+19}_{-15}$ & $59^{+207}_{-32}$ \\
(148780) Altjira & 0.003 & $3.931^{+0.058}_{-0.057}$ & $9886^{+50}_{-49}$ & $0.3433^{+0.0041}_{-0.0040}$ & $25.195^{+0.204}_{-0.206}$ & $192.89^{+0.67}_{-0.66}$ & $274.59^{+0.36}_{-0.36}$ & $122.76^{+0.76}_{-0.75}$ \\
(160091) 2000 OL$_{67}$, \textit{pro.} & 0.047 & $0.289^{+0.026}_{-0.023}$ & $7600^{+224}_{-210}$ & $0.2279^{+0.0211}_{-0.0211}$ & $87.2^{+5.9}_{-5.4}$ & $37.42^{+3.51}_{-3.14}$ & $226.48^{+4.46}_{-3.75}$ & $87.08^{+2.74}_{-3.13}$ \\
(160091) 2000 OL$_{67}$, \textit{retro.} & 0.045 & $0.315^{+0.033}_{-0.027}$ & $7816^{+262}_{-230}$ & $0.2226^{+0.0271}_{-0.0253}$ & $88.0^{+10.8}_{-10.7}$ & $39.78^{+4.64}_{-3.91}$ & $290.35^{+3.69}_{-3.65}$ & $85.71^{+3.50}_{-3.82}$ \\
(160256) 2002 PD$_{149}$ & 0.011 & $0.540^{+0.022}_{-0.020}$ & $26748^{+333}_{-320}$ & $0.5886^{+0.0091}_{-0.0089}$ & $26.760^{+0.487}_{-0.476}$ & $144.98^{+1.18}_{-1.19}$ & $96.47^{+1.11}_{-1.11}$ & $171.41^{+2.64}_{-2.63}$ \\
(174567) Varda-Ilmarë, \textit{pro.} & 0.005 & $265.150^{+5.379}_{-5.435}$ & $4801^{+32}_{-33}$ & $0.0240^{+0.0055}_{-0.0053}$ & $68.323^{+1.975}_{-1.964}$ & $329.3^{+12.6}_{-14.6}$ & $315.95^{+1.81}_{-1.67}$ & $137.9^{+14.3}_{-12.8}$ \\
(174567) Varda-Ilmarë, \textit{retro.} & 0.005 & $265.399^{+5.748}_{-5.256}$ & $4802^{+34}_{-32}$ & $0.0174^{+0.0048}_{-0.0049}$ & $77.217^{+1.967}_{-2.064}$ & $298^{+19}_{-22}$ & $2.92^{+1.87}_{-1.60}$ & $154^{+22}_{-19}$ \\
(225088) Gonggong-Xiangliu, \textit{pro.} & 0.006 & $1765.784^{+70.143}_{-71.451}$ & $24201^{+316}_{-331}$ & $0.2852^{+0.0086}_{-0.0079}$ & $63.652^{+1.461}_{-1.404}$ & $63.14^{+1.50}_{-1.63}$ & $35.11^{+1.55}_{-1.59}$ & $97.31^{+1.87}_{-1.85}$ \\
(225088) Gonggong-Xiangliu, \textit{retro.} & 0.005 & $1728.671^{+59.377}_{-52.122}$ & $24032^{+272}_{-244}$ & $0.2910^{+0.0085}_{-0.0079}$ & $121.456^{+1.358}_{-1.308}$ & $67.04^{+1.59}_{-1.55}$ & $89.17^{+1.73}_{-1.83}$ & $95.54^{+2.05}_{-2.03}$ \\
(229762) G!kún\textdoublepipe’hòmdímà-G!ò’é !hú & 0.016 & $136.625^{+3.682}_{-3.883}$ & $6044^{+54}_{-58}$ & $0.0234^{+0.0079}_{-0.0064}$ & $55.444^{+0.735}_{-0.748}$ & $312^{+16}_{-24}$ & $127.99^{+1.15}_{-1.10}$ & $254^{+23}_{-15}$ \\
(275809) 2001 QY$_{297}$ & 0.0004 & $4.109^{+0.038}_{-0.038}$ & $9964^{+31}_{-30}$ & $0.4174^{+0.0023}_{-0.0023}$ & $161.001^{+0.107}_{-0.108}$ & $316.84^{+0.49}_{-0.49}$ & $15.47^{+0.74}_{-0.72}$ & $181.45^{+0.43}_{-0.44}$ \\
(341520) Mors-Somnus & 0.023 & $0.745^{+0.020}_{-0.020}$ & $20657^{+154}_{-153}$ & $0.1568^{+0.0023}_{-0.0023}$ & $15.62^{+0.189}_{-0.188}$ & $320.15^{+2.72}_{-2.81}$ & $331.42^{+0.92}_{-0.95}$ & $294.52^{+3.62}_{-3.54}$ \\
(364171) 2006 JZ$_{81}$ & 0.047 & $1.332^{+0.236}_{-0.198}$ & $33338^{+1755}_{-1653}$ & $0.8517^{+0.0100}_{-0.0112}$ & $12.631^{+0.676}_{-0.574}$ & $170.64^{+0.79}_{-0.78}$ & $80.62^{+2.42}_{-2.22}$ & $75.41^{+1.91}_{-2.16}$ \\ 
(385446) Manwë-Thorondor & 0.011 & $1.946^{+0.035}_{-0.034}$ & $6681^{+40}_{-39}$ & $0.5643^{+0.0067}_{-0.0065}$ & $48.541^{+0.252}_{-0.245}$ & $77.81^{+1.93}_{-1.96}$ & $170.62^{+0.51}_{-0.51}$ & $236.50^{+1.98}_{-1.95}$ \\
(469705) \kagara{} & 0.010 & $2.167^{+0.131}_{-0.119}$ & $7656^{+151}_{-142}$ & $0.6928^{+0.0137}_{-0.0135}$ & $10.150^{+0.412}_{-0.412}$ & $170.21^{+0.94}_{-0.98}$ & $12.84^{+1.07}_{-1.02}$ & $312.57^{+0.42}_{-0.39}$ \\
(508788) 2000 CQ$_{114}$ & 0.001 & $0.544^{+0.008}_{-0.008}$ & $6935^{+35}_{-34}$ & $0.096^{+0.0047}_{-0.0045}$ & $44.145^{+0.461}_{-0.464}$ & $333.60^{+3.75}_{-3.89}$ & $132.92^{+0.70}_{-0.72}$ & $309.21^{+4.54}_{-4.43}$ \\
(508869) 2002 VT$_{130}$, \textit{pro.} & 0.002 & $2.325^{+0.211}_{-0.196}$ & $3028^{+89}_{-87}$ & $0.0207^{+0.0089}_{-0.0082}$ & $65.465^{+3.257}_{-2.755}$ & $48^{+299}_{-37}$ & $127.92^{+2.33}_{-2.31}$ & $294^{+28}_{-39}$ \\
(508869) 2002 VT$_{130}$, \textit{retro.} & 0.002 & $2.255^{+0.153}_{-0.145}$ & $2998^{+66}_{-66}$ & $0.0219^{+0.0073}_{-0.0069}$ & $111.021^{+2.707}_{-2.898}$ & $73^{+29}_{-31}$ & $191.87^{+2.18}_{-2.22}$ & $231^{+31}_{-28}$ \\
(524366) 2001 XR$_{254}$ & 0.002 & $4.044^{+0.062}_{-0.060}$ & $9302^{+49}_{-48}$ & $0.5558^{+0.0046}_{-0.0045}$ & $20.124^{+0.203}_{-0.202}$ & $286.95^{+0.86}_{-0.86}$ & $321.94^{+0.49}_{-0.48}$ & $265.71^{+0.68}_{-0.67}$ \\
(524531) 2002 XH$_{91}$ & 0.009 & $6.469^{+0.415}_{-0.354}$ & $22405^{+463}_{-410}$ & $0.7125^{+0.0155}_{-0.0162}$ & $33.865^{+1.407}_{-1.545}$ & $196.61^{+1.82}_{-1.49}$ & $239.11^{+2.14}_{-2.23}$ & $203.92^{+2.43}_{-2.98}$ \\
(525462) 2005 EO$_{304}$ & 0.054 & $2.131^{+0.023}_{-0.023}$ & $69601^{+226}_{-223}$ & $0.2154^{+0.0012}_{-0.0012}$ & $13.283^{+0.071}_{-0.071}$ & $208.68^{+2.15}_{-2.11}$ & $257.94^{+0.93}_{-0.94}$ & $185.32^{+1.19}_{-1.20}$ \\
(612095) 1999 OJ$_{4}$ & 0.010 & $0.406^{+0.007}_{-0.007}$ & $3310^{+19}_{-19}$ & $0.3675^{+0.0039}_{-0.0038}$ & $54.460^{+0.819}_{-0.820}$ & $82.44^{+1.09}_{-1.09}$ & $259.39^{+1.19}_{-1.22}$ & $67.85^{+0.84}_{-0.83}$ \\
(612147) 2000 CF$_{105}$ & 0.050 & $0.187^{+0.003}_{-0.003}$ & $33387^{+119}_{-117}$ & $0.2978^{+0.0080}_{-0.0079}$ & $167.585^{+0.177}_{-0.180}$ & $296.97^{+1.27}_{-1.25}$ & $224.86^{+0.85}_{-0.85}$ & $262.36^{+0.81}_{-0.83}$ \\
(612176) 2000 QL$_{251}$ & 0.003 & $3.094^{+0.028}_{-0.028}$ & $4992^{+15}_{-15}$ & $0.4897^{+0.0040}_{-0.0040}$ & $131.653^{+0.570}_{-0.568}$ & $12.28^{+0.32}_{-0.32}$ & $89.64^{+0.90}_{-0.90}$ & $353.91^{+0.63}_{-0.62}$ \\
(612239) 2001 QC$_{298}$ & 0.002 & $11.853^{+0.154}_{-0.143}$ & $3810^{+16}_{-15}$ & $0.3342^{+0.0024}_{-0.0024}$ & $54.994^{+0.384}_{-0.382}$ & $209.65^{+0.27}_{-0.27}$ & $95.18^{+0.51}_{-0.49}$ & $353.13^{+0.43}_{-0.42}$ \\
(612687) 2003 UN$_{284}$ & 0.076 & $1.270^{+0.084}_{-0.076}$ & $53235^{+1068}_{-1006}$ & $0.375^{+0.0183}_{-0.0178}$ & $25.671^{+0.304}_{-0.308}$ & $168.52^{+1.22}_{-1.14}$ & $93.48^{+1.17}_{-1.12}$ & $291.05^{+2.45}_{-2.57}$ \\
1998 WW$_{31}$ & 0.010 & $2.663^{+0.016}_{-0.016}$ & $22631^{+46}_{-46}$ & $0.8187^{+0.0021}_{-0.0021}$ & $48.055^{+0.396}_{-0.385}$ & $221.74^{+0.38}_{-0.38}$ & $117.79^{+0.60}_{-0.58}$ & $42.35^{+0.35}_{-0.35}$ \\
1999 RT$_{214}$ & 0.004 & $0.192^{+0.010}_{-0.009}$ & $3386^{+56}_{-55}$ & $0.2904^{+0.0217}_{-0.0219}$ & $24.112^{+1.309}_{-1.243}$ & $77.65^{+2.97}_{-2.99}$ & $74.21^{+2.65}_{-2.70}$ & $329.08^{+2.87}_{-2.91}$ \\
2001 QW$_{322}$ & 0.092 & $2.032^{+0.023}_{-0.022}$ & $103657^{+317}_{-310}$ & $0.4592^{+0.0015}_{-0.0015}$ & $150.789^{+0.172}_{-0.170}$ & $251.31^{+0.25}_{-0.25}$ & $241.52^{+0.29}_{-0.29}$ & $164.68^{+0.26}_{-0.26}$ \\
2003 QY$_{90}$ & 0.016 & $0.516^{+0.017}_{-0.015}$ & $8544^{+90}_{-84}$ & $0.6602^{+0.0103}_{-0.0098}$ & $54.581^{+0.463}_{-0.441}$ & $146.53^{+0.89}_{-0.87}$ & $131.58^{+0.48}_{-0.46}$ & $26.00^{+2.16}_{-2.28}$ \\
2003 TJ$_{58}$ & 0.002 & $0.237^{+0.010}_{-0.009}$ & $3842^{+51}_{-48}$ & $0.5169^{+0.0081}_{-0.0079}$ & $61.650^{+1.803}_{-1.762}$ & $254.62^{+1.92}_{-2.01}$ & $193.33^{+2.34}_{-2.42}$ & $333.33^{+1.89}_{-1.85}$ \\
2004 PB$_{108}$ & 0.001 & $9.443^{+0.325}_{-0.312}$ & $10389^{+121}_{-119}$ & $0.4368^{+0.0065}_{-0.0065}$ & $101.278^{+0.956}_{-0.929}$ & $209.80^{+0.42}_{-0.41}$ & $120.01^{+0.81}_{-0.84}$ & $176.59^{+0.33}_{-0.33}$ \\
2006 BR$_{284}$ & 0.040 & $0.580^{+0.007}_{-0.007}$ & $25454^{+106}_{-104}$ & $0.2746^{+0.0018}_{-0.0018}$ & $55.019^{+0.596}_{-0.584}$ & $13.41^{+0.52}_{-0.51}$ & $42.47^{+0.46}_{-0.46}$ & $182.18^{+0.57}_{-0.57}$ \\
2006 CH$_{69}$ & 0.046 & $1.223^{+0.066}_{-0.067}$ & $31485^{+567}_{-591}$ & $0.9174^{+0.0022}_{-0.0023}$ & $132.742^{+0.802}_{-0.789}$ & $154.80^{+0.64}_{-0.69}$ & $111.09^{+0.82}_{-0.87}$ & $234.83^{+0.86}_{-0.80}$ \\
\enddata
\tablecomments{All fitted angles are relative to the J2000 ecliptic plane. Listed errors are 16 and 84 percentile levels. }
\label{tab:keplerian}
\end{deluxetable}
\endlongrotatetable

\longrotatetable
\begin{deluxetable}{lcccccccccccc}
\tabletypesize{\footnotesize}
\tablewidth{\textwidth}
\tablecaption{Non-Keplerian Orbital Fits}
\tablewidth{0pt}
\tablehead{
Object & $\frac{\mathcal{L}_K}{\mathcal{L}_{NK}}$ & Residual & $M_{sys}$ & $a$ & $e$ & $i$ & $\omega$ & $\Omega$ & $\mathcal{M}$ & $J_2$ & $i_{sp}$ & $\Omega_{sp}$ \\
 & & (arcsec) & ($10^{18}$ kg) & (km) & & ($^{\circ}$) & ($^{\circ}$) & ($^{\circ}$) & ($^{\circ}$) & & ($^{\circ}$) & ($^{\circ}$) 
}
\startdata
(26308) 1998 SM$_{165}$ & $0.18142$ & 0.002 & $6.884$ & $11374$ & $0.4733$ & $66.296$ & $49.286$ & $52.253$ & $165.594$ & $0.3425$ & $104.833$ & $67.125$ \\
(42355) Typhon-Echidna & $0.33001$ & 0.0004 & $1.105$ & $1717$ & $0.5178$ & $43.748$ & $124.229$ & $235.400$ & $24.487$ & $0.1291$ & $25.870$ & $197.303$ \\
(50000) Quaoar-Weywot, \textit{pro.} & $0.53267$ & 0.003 & $1211.280$ & $13314$ & $0.0209$ & $13.880$ & $299.937$ & $354.925$ & $233.712$ & $0.0604$ & $120.289$ & $55.900$ \\
(50000) Quaoar-Weywot, \textit{retro.} & $0.54467$ & 0.004 & $1214.346$ & $13328$ & $0.0249$ & $150.171$ & $298.836$ & $345.454$ & $240.463$ & $0.0228$ & $157.314$ & $118.475$ \\
(55637) 2002 UX$_{25}$ & $0.65209$ & 0.009 & $123.885$ & $4761$ & $0.1754$ & $84.330$ & $17.006$ & $203.272$ & $288.844$ & $0.0040$ & $91.551$ & $224.767$ \\
(58534) Logos-Zoe & $0.95946$ & 0.005 & $0.450$ & $8166$ & $0.5396$ & $72.750$ & $201.645$ & $17.141$ & $35.030$ & $0.4268$ & $124.195$ & $45.927$ \\
(60458) 2000 CM$_{114}$ & $0.10142$ & 0.003 & $1.931$ & $2468$ & $0.0399$ & $74.361$ & $203.330$ & $270.567$ & $15.880$ & $0.2368$ & $128.480$ & $20.058$ \\
(65489) Ceto-Phorcus, \textit{pro.} & $0.70688$ & 0.001 & $5.323$ & $1831$ & $0.0012$ & $77.513$ & $107.102$ & $111.404$ & $147.204$ & $0.0508$ & $103.094$ & $316.276$ \\
(65489) Ceto-Phorcus, \textit{retro.} & $0.17520$ & 0.002 & $5.529$ & $1852$ & $0.0026$ & $132.016$ & $41.156$ & $122.304$ & $215.553$ & $0.1403$ & $160.518$ & $337.142$ \\
\textbf{(66652) Borasisi-Pabu} & \textbf{0.01465} & \textbf{0.002} & \textbf{3.445} & \textbf{4531} & \textbf{0.4680} & \textbf{49.715} & \textbf{139.836} & \textbf{86.232} & \textbf{335.745} & \textbf{0.2789} & \textbf{95.761} & \textbf{79.983} \\
\textbf{(79360) Sila-Nunam} & \textbf{0.04804} & \textbf{0.003} & \textbf{10.951} & \textbf{2788} & \textbf{0.0168} & \textbf{120.243} & \textbf{150.972} & \textbf{134.927} & \textbf{68.092} & \textbf{0.1989} & \textbf{134.511} & \textbf{104.933} \\
(80806) 2000 CM$_{105}$ & $1.02954$ & 0.006 & $0.834$ & $2358$ & $0.0139$ & $61.749$ & $243.138$ & $201.089$ & $288.276$ & $0.0021$ & $102.200$ & $54.812$ \\
(88611) Teharonhiawako-Sawiskera & $0.11301$ & 0.027 & $2.424$ & $27564$ & $0.2476$ & $126.786$ & $302.060$ & $36.055$ & $275.907$ & $1.7920$ & $127.862$ & $322.045$ \\
\textbf{(90482) Orcus-Vanth} & \textbf{0.00170} & \textbf{0.003} & \textbf{634.265} & \textbf{8995} & \textbf{0.0008} & \textbf{90.484} & \textbf{285.772} & \textbf{50.549} & \textbf{190.674} & \textbf{0.0158} & \textbf{103.357} & \textbf{118.930} \\
(119979) 2002 WC$_{19}$, \textit{pro.} & $0.84427$ & 0.010 & $79.218$ & $4132$ & $0.2208$ & $23.712$ & $136.709$ & $121.284$ & $359.573$ & $0.0143$ & $58.997$ & $160.318$ \\
(119979) 2002 WC$_{19}$, \textit{retro.} & $0.21941$ & 0.009 & $73.095$ & $4023$ & $0.2419$ & $162.290$ & $169.630$ & $243.760$ & $343.458$ & $0.0235$ & $157.775$ & $85.412$ \\
\textbf{(120347) Salacia-Actaea} & \textbf{0.00028} & \textbf{0.011} & \textbf{498.199} & \textbf{5746} & \textbf{0.0052} & \textbf{17.512} & \textbf{276.631} & \textbf{110.476} & \textbf{273.107} & \textbf{0.0195} & \textbf{109.417} & \textbf{306.670} \\
(123509) 2000 WK$_{183}$, \textit{blended} & $0.40756$ & 0.003 & $1.130$ & $2388$ & $0.0114$ & $98.754$ & $168.892$ & $151.805$ & $296.359$ & $0.4792$ & $109.633$ & $275.744$ \\
(134860) 2000 OJ$_{67}$, \textit{blended} & $0.96827$ & 0.002 & $1.877$ & $2257$ & $0.0358$ & $81.435$ & $151.329$ & $257.076$ & $27.989$ & $0.0145$ & $81.781$ & $12.101$ \\
\textbf{(148780) Altjira} & \textbf{0.01624} & \textbf{0.001} & \textbf{4.003} & \textbf{9950} & \textbf{0.3474} & \textbf{25.176} & \textbf{191.938} & \textbf{274.506} & \textbf{123.690} & \textbf{0.4787} & \textbf{159.051} & \textbf{187.387}\\
(160091) 2000 OL$_{67}$, \textit{pro.} & $0.99577$ & 0.005 & $0.2850$ & $7563$ & $0.232$ & $85.499$ & $37.090$ & $226.364$ & $87.724$ & $0.0311$ & $86.771$ & $291.541$ \\
(160091) 2000 OL$_{67}$, \textit{retro.} & $1.01109$ & 0.004 & $0.308$ & $7757$ & $0.2169$ & $85.271$ & $40.593$ & $288.921$ & $84.932$ & $0.0298$ & $114.926$ & $248.907$ \\
(160256) 2002 PD$_{149}$ & $0.81370$ & 0.008 & $0.550$ & $26907$ & $0.5907$ & $26.377$ & $143.969$ & $96.259$ & $172.878$ & $1.0548$ & $67.386$ & $97.583$ \\
\textbf{(174567) Varda-Ilmarë, \textit{pro.}} & \textbf{0.02790} & \textbf{0.004} & \textbf{266.803} & \textbf{4808} & \textbf{0.0119} & \textbf{63.392} & \textbf{296.882} & \textbf{313.943} & \textbf{171.285} & \textbf{0.0702} & \textbf{127.958} & \textbf{297.209} \\
(174567) Varda-Ilmarë, \textit{retro.} & $0.13531$ & 0.005 & $267.863$ & $4817$ & $0.0178$ & $79.784$ & $309.956$ & $2.040$ & $142.580$ & $0.2404$ & $115.840$ & $274.627$ \\
\textbf{(225088) Gonggong-Xiangliu, \textit{pro.}} & \textbf{0.00620} & \textbf{0.003} & \textbf{1767.826} & \textbf{24206} & \textbf{0.2988} & \textbf{61.563} & \textbf{65.191} & \textbf{38.064} & \textbf{94.381} & \textbf{0.0841} & \textbf{108.045} & \textbf{73.731} \\
(225088) Gonggong-Xiangliu, \textit{retro.} & $0.24885$ & 0.003 & $1758.133$ & $24163$ & $0.2958$ & $121.773$ & $67.608$ & $88.035$ & $95.070$ & $0.1414$ & $167.678$ & $265.320$ \\
(229762) G!kún\textdoublepipe’hòmdímà-G!ò’é !hú & $0.59484$ & 0.016 & $136.330$ & $6038$ & $0.0221$ & $55.209$ & $308.304$ & $127.772$ & $258.865$ & $0.0109$ & $82.470$ & $289.514$ \\
(275809) 2001 QY$_{297}$ & $1.01557$ & 0.0004 & $4.113$ & $9965$ & $0.4174$ & $160.987$ & $316.890$ & $15.553$ & $181.453$ & $0.0018$ & $161.279$ & $68.401$ \\
(341520) Mors-Somnus & $0.97259$ & 0.023 & $0.745$ & $20644$ & $0.1568$ & $15.623$ & $320.190$ & $331.366$ & $294.476$ & $0.1022$ & $155.097$ & $235.952$ \\
(364171) 2006 JZ$_{81}$ & $0.95893$ & 0.047 & $1.291$ & $33008$ & $0.8478$ & $12.746$ & $170.621$ & $81.052$ & $75.688$ & $0.4006$ & $162.322$ & $333.048$ \\
(385446) Manwë-Thorondor & $0.66357$ & 0.011 & $1.932$ & $6663$ & $0.5646$ & $48.480$ & $78.514$ & $170.743$ & $235.848$ & $0.2750$ & $77.739$ & $92.703$ \\
(469705) \kagara{} & $0.73924$ & 0.010 & $2.110$ & $7587$ & $0.6885$ & $10.272$ & $170.258$ & $13.072$ & $312.525$ & $0.2921$ & $162.551$ & $82.769$ \\
(508788) 2000 CQ$_{114}$ & $0.95243$ & 0.001 & $0.542$ & $6928$ & $0.0963$ & $44.124$ & $334.453$ & $132.858$ & $308.338$ & $0.3054$ & $89.798$ & $333.487$ \\
(508869) 2002 VT$_{130}$, \textit{pro.} & $0.60596$ & 0.002 & $2.313$ & $3023$ & $0.0144$ & $65.104$ & $33.765$ & $129.898$ & $272.126$ & $0.1159$ & $103.309$ & $85.782$ \\
(508869) 2002 VT$_{130}$, \textit{retro.} & $0.25955$ & 0.002 & $2.381$ & $3052$ & $0.0226$ & $111.519$ & $9.521$ & $197.281$ & $296.199$ & $0.2488$ & $105.822$ & $119.414$ \\
(524366) 2001 XR$_{254}$ & $0.86142$ & 0.002 & $4.071$ & $9323$ & $0.5547$ & $20.099$ & $287.158$ & $321.814$ & $265.548$ & $0.3499$ & $166.597$ & $11.987$ \\
(524531) 2002 XH$_{91}$ & $0.86499$ & 0.008 & $7.519$ & $23567$ & $0.7373$ & $32.289$ & $195.870$ & $242.551$ & $203.976$ & $0.4515$ & $125.906$ & $90.043$ \\
(525462) 2005 EO$_{304}$ & $0.20246$ & 0.054 & $2.125$ & $69573$ & $0.2148$ & $13.212$ & $209.267$ & $258.120$ & $184.769$ & $2.7635$ & $142.351$ & $296.259$ \\
(612095) 1999 OJ$_{4}$ & $0.93881$ & 0.010 & $0.406$ & $3309$ & $0.3671$ & $54.519$ & $82.718$ & $259.123$ & $67.569$ & $0.2666$ & $64.583$ & $140.452$ \\
(612147) 2000 CF$_{105}$ & $0.86019$ & 0.050 & $0.186$ & $33379$ & $0.2952$ & $167.561$ & $296.256$ & $224.185$ & $262.300$ & $3.8540$ & $127.820$ & $221.565$ \\
(612176) 2000 QL$_{251}$ & $0.77542$ & 0.002 & $3.111$ & $4997$ & $0.4885$ & $132.000$ & $12.407$ & $89.402$ & $353.609$ & $0.2289$ & $102.649$ & $151.539$ \\
(612239) 2001 QC$_{298}$ & $0.37771$ & 0.002 & $11.924$ & $3818$ & $0.3357$ & $54.928$ & $207.807$ & $96.128$ & $355.324$ & $0.0592$ & $30.340$ & $104.721$ \\
(612687) 2003 UN$_{284}$ & $1.01884$ & 0.076 & $1.273$ & $53264$ & $0.3736$ & $25.629$ & $168.468$ & $93.554$ & $290.990$ & $0.2812$ & $96.227$ & $7.402$ \\
1998 WW$_{31}$ & $0.64207$ & 0.008 & $2.682$ & $22668$ & $0.8164$ & $48.109$ & $221.621$ & $116.717$ & $42.893$ & $0.4753$ & $99.194$ & $347.466$ \\
\textbf{1999 RT$_{214}$} & \textbf{0.03836} & \textbf{0.003} & \textbf{0.202} & \textbf{3441} & \textbf{0.2893} & \textbf{24.819} & \textbf{71.483} & \textbf{79.721} & \textbf{331.771} & \textbf{0.4357} & \textbf{131.628} & \textbf{180.017} \\
2001 QW$_{322}$ & $0.99997$ & 0.092 & $2.033$ & $103653$ & $0.4592$ & $150.778$ & $251.286$ & $241.500$ & $164.669$ & $0.0538$ & $114.421$ & $153.147$ \\
2003 QY$_{90}$ & $0.96631$ & 0.016 & $0.516$ & $8542$ & $0.6596$ & $54.590$ & $146.471$ & $131.557$ & $26.008$ & $0.3543$ & $125.041$ & $151.681$ \\
2003 TJ$_{58}$ & $0.88836$ & 0.002 & $0.240$ & $3856$ & $0.5186$ & $61.075$ & $254.512$ & $193.323$ & $333.211$ & $0.4378$ & $149.733$ & $168.308$ \\
2004 PB$_{108}$ & $0.95583$ & 0.001 & $9.379$ & $10363$ & $0.4346$ & $101.491$ & $209.943$ & $119.949$ & $176.609$ & $0.4518$ & $138.515$ & $297.249$ \\
2006 BR$_{284}$ & $0.10005$ & 0.040 & $0.576$ & $25359$ & $0.2760$ & $55.354$ & $13.799$ & $41.542$ & $182.234$ & $3.8250$ & $139.757$ & $54.153$ \\
2006 CH$_{69}$ & $0.77312$ & 0.045 & $1.178$ & $31089$ & $0.9145$ & $132.054$ & $154.503$ & $110.849$ & $234.770$ & $0.4294$ & $128.997$ & $96.168$ \\
\enddata
\tablecomments{All fitted angles are relative to the J2000 ecliptic plane. Residual indicates the combined RMS residual for the best fit. In our fits, we use $\ln{J_2R^2}$ as our fitted parameter. Here, we present only the $J_2$ to enable comparison with the literature. For $R$, we use the volumetric radius of the modeled body (see Table \ref{tab:tnbs}). Bold-face data indicates TNBs that have detections of non-Keplerian motion.}
\label{tab:nonkeplerian}
\end{deluxetable}
\endlongrotatetable

\bibliographystyle{apj}
\bibliography{all}

\end{document}